\documentclass[journal]{IEEEtran}

\usepackage{amsmath}
\usepackage{amssymb}
\usepackage{floatflt,epsfig}
\usepackage{algorithm}
\usepackage{algorithmic}
\usepackage{inputenc}
\usepackage{graphics}
\usepackage{graphicx}
\usepackage{epic}
\usepackage{eepic}
\usepackage{psfrag}
\usepackage{amsfonts}
\usepackage{array}
\usepackage{tikz}
\usetikzlibrary{calc,backgrounds,arrows,matrix,shapes,automata,petri}

\usepackage{stfloats}
\usepackage{epsfig}
\usepackage{psfrag}
\usepackage[caption=false,font=footnotesize]{subfig}
\usepackage{color}
\usepackage{amssymb}
\usepackage{array}
\usepackage{mathtools}
\usepackage{wrapfig}
\usepackage{cite}

\newtheorem{example}{Example}

\begin{document}

\title{Buffer-Based Distributed LT Codes}

\author{
Iqbal Hussain,~\IEEEmembership{Student Member,~IEEE,} Ming Xiao,~\IEEEmembership{Senior Member,~IEEE,}\\
and Lars K. Rasmussen,~\IEEEmembership{Senior Member,~IEEE}%
\thanks{The authors are with the School of Electrical Engineering and the ACCESS Linnaeus Center, KTH Royal Institute of Technology, Stockholm, Sweden. E-mail: \{iqbalh,mingx,lkra\}@kth.se. L. K. Rasmussen is also affiliated with the Institute for Telecommunication Research, University of South Australia, Adelaide, Australia, in an adjunct position.} %
\thanks{The research leading to these results has received funding from the Swedish Research Council under VR grants 621-2009-4666, 621-2008-4249.}%
}
\maketitle

\vspace{-10mm}

\begin{abstract}
We focus on the design of distributed Luby transform (DLT) codes for erasure networks with multiple sources and multiple relays, communicating to a single destination. The erasure-floor performance of DLT codes improves with the maximum degree of the relay-degree distribution. However, for conventional DLT codes, the maximum degree is upper-bounded by the number of sources. An additional constraint is that the sources are required to have the same information block length. We introduce a $D$-bit buffer for each source-relay link, which allows the relay to select multiple encoded bits from the same source for the relay-encoding process; thus, the number of sources no longer limits the maximum degree at the relay. Furthermore, the introduction of buffers facilitates the use of different information block sizes across sources. Based on density evolution we develop an asymptotic analytical framework for optimization of the relay-degree distribution. We further integrate techniques for unequal erasure protection into the optimization framework. The proposed codes are considered for both lossless and lossy source-relay links. Numerical examples show that there is no loss in erasure performance for transmission over lossy source-relay links as compared to lossless links. Additional delays, however, may occur. The design framework and our contributions are demonstrated by a number of illustrative examples, showing the improvements obtained by the proposed buffer-based DLT codes.
\end{abstract}

\section{Introduction}

The fountain-code concept was suggested in \cite{BLM02} for reliable broadcast/multicast in packet-based transmission at higher-layers in wired and wireless networks. At higher layers the link model is typically assumed to be a packet erasure channel (PEC) with unknown erasure probabilities. A fountain code is inherently rateless, and as a consequence such codes may potentially produce an unlimited number of encoded bits from a given limited block of information bits in order to adapt to the link capacity. Hence, in contrast to fixed-rate codes, the code rate of rateless codes is not fixed prior to transmission but adapted on-the-fly. Fountain codes are therefore ideally suited for file \cite{GR05} and multimedia \cite{MagWanFroMar13TMM,VukKhiStaTho14TMM} distribution in networks with unknown link qualities.

The Luby transform (LT) code was the first practical realization of rateless codes \cite{LT}, adapting universally to any unknown link erasure probability as the information block length grows large. However, each instance of an LT code is a low-density generator-matrix (LDGM) code, and thus the minimum distance of the code is inherently poor, leading to a high erasure floor \cite{LDGM2003}. To solve this problem Raptor codes were proposed, where a high-rate precode is introduced \cite{Sho06}. In addition to alleviating the erasure floor, Raptor codes also provide lower decoding complexity. The design and performance of both LT codes and Raptor codes have also been investigated for physical-layer models, such as additive white Gaussian noise (AWGN) channels \cite{PalYed04isit, HusXiaRas11globe, MFY14}, Rayleigh fading channels \cite{CasMao06CL}, and relay channels \cite{CasMao07TWC}. A survey of developments in rateless coding can be found in \cite{MFY14-2}.

Extensions to multi-source, multi-relay networks with erasure links is desirable, since multicast transmission involving relaying is an emerging technology in current and upcoming cellular standards \cite{VukKhiStaTho14TMM}. Multi-relay network is in particular considered as a promising approach for improving coverage and increasing throughput for future broadband communication networks. Additional motivation is provided by emerging applications of large-scale wireless sensor networks \cite{DimPraRam06IT}, as demonstrated by standards activities in IEEE 802.11, 802.15, and 802.16j. A large number of sensor nodes may gather information from the surrounding environment, which is subsequently forwarded to a common destination through a set of dedicated relaying sensor nodes.

For such extensions the question arises whether to consider random linear network coding or network coding based on distributed LT (DLT) codes \cite{MagWanFroMar13TMM,VukKhiStaTho14TMM}. The choice is between very low-complexity encoding and high-complexity decoding, characterizing random linear network coding; and moderate-complexity encoding and low-complexity decoding, characterizing LT codes. As random linear network coding has been extensively investigated in the literature, we focus primarily on DLT codes. The first DLT codes were introduced in \cite{PKF06,PKF07} for two and four sources, communicating to a single destination via a relay. The degree distribution at the sources and the combining operation at the relay are coordinated to obtain a Soliton-like check-node degree distribution\footnote{The ideal check-node degree distribution for an LT code is the ideal Soliton distribution; however, this distribution is quite fragile to statistical variations. Instead the Robust Soliton distribution was proposed in \cite{LT}, ensuring successful decoding with high probability.}, thus realizing an equivalent conventional LT code, at the destination. A similar approach was suggested in \cite{GumSre08itw}, while in \cite{ChaHugKerSco10icdcs} the sources use the Robust Soliton distribution (RSD), and the relay combining is modified to ensure a resulting RSD at the destination. The ideas are extended to cater for multiple destinations with direct source-destination links in \cite{SuhHeo11vtc}. However, in all cases the complexity of the combining operation at the relay is relatively high. To overcome this problem, a more general approach to DLT codes was investigated in \cite{SejVukDouSenPie07acssc,SPD09,SPDI10}, where the relay independently selects the incoming packets from the sources. A corresponding design framework based on AND-OR tree analysis was formulated for DLT codes over a binary erasure channel (BEC). As the number of sources limits the maximum degree of the relay-degree distribution, the number of sources likewise determines the erasure floor.
A relay combining algorithm was proposed in \cite{LYK11} for a Y-network such that received symbols at the destination follow a Soliton-like distribution without a specific design of the degree distribution at each source. The scheme in \cite{LYK11} outperforms the strategies presented in \cite{PKF07,SPD09} for a two-source scenario. However, the approach requires a large memory buffer for each source-relay link at the relay. Moreover, the scheme is not able to ensure an optimal degree distribution at the relay, and thus only a Soliton-like distribution is realized at the destination. In \cite{LiauKimYousefi13TCOM} the approach is improved by performing distribution-shaping at the relay to ensure more effective message-passing decoding at the destination. 

DLT codes have also been considered for physical-layer wireless channels. In \cite{PangLinUchoaVucetic12WCL}, a DLT code is realized for a multi-source, multi-relay, single-destination network, where the sources transmit uncoded packets to the relays; in turn, the relays transmit LT-coded packets using the RSD as the degree distribution. Analytical upper bounds are determined for the symbol-error rate (SER). The approach is further developed in \cite{YueLinVucMaoAul13TCOM}, where lower-bounds are derived on the SER for the DLT scheme in \cite{PangLinUchoaVucetic12WCL}. Also, the DLT scheme is extended to a distributed Raptor (DR) code, where a number of relay nodes are selected with a given probability to perform pre-coding. Upper and lower bounds on the resulting SER are derived. A similar approach is proposed in \cite{Sugiura12CL}, where a DR code is realized by randomly selecting source nodes to perform pre-coding. Performance closely resembling a stand-alone Raptor code is obtained.

Equal erasure protection (EEP) is an important property of many transmission schemes. However, in certain applications, unequal erasure protection (UEP) of data is required. For example, in wireless packetized networks, the header may have a higher priority than the payload. Similarly in video transmission, the picture header and motion vectors are more important than the texture data. At the physical layer, UEP is obtained by forward-error-control codes and hierarchical modulation, where fixed-rate codes provide additional redundancy for higher-priority data to realize UEP. However, in scenarios where the packet erasure probabilities vary, fixed-rate coding suffers a number of drawbacks \cite{ASL07}. Therefore, rateless codes that adapt to changing link qualities become a natural choice for enabling UEP capabilities for packet transmission. This was first considered in \cite{RVF07} for point-to-point transmission, where classes of higher priority bits are used more frequently in the encoding process by weighting the selection probability. Optimized degree distributions were proposed in \cite{HusXiaRas13CL} to significantly improve the erasure floor. An alternative approach, based on an expanding encoding window, was proposed in \cite{SejVukDouSenPie07acssc,SVDSP09}.

Due to the presence of a relay node and its combining process, it is a challenging task to realize UEP for DLT codes. A first attempt was made in \cite{TR12}, extending the weighted UEP approach in \cite{RVF07}. Here, distributed LT codes were designed and optimized for UEP using a multi-objective genetic algorithm. However, due to the complexity of multi-parameters optimization, the scheme in \cite{TR12} is restricted to only two sources, and yet to be extended. In \cite{ShaXuZha13CL} the approach in \cite{SejVukDouSenPie07acssc,SPD09,SPDI10} is extended to enable UEP, using the weighted strategy in \cite{RVF07}. Source and relay degree distributions are optimized recursively through linear and non-linear programming, demonstrating improved performance. Extensions to UEP schemes for physical-layer wireless channels have been considered in \cite{YueLinLiBaiVuc13wcnc}, introducing the weighted selection strategy into the scenario considered in \cite{PangLinUchoaVucetic12WCL}. Upper and lower bounds are obtained for maximum-likelihood (ML) decoding over a Rayleigh fading channel. The approach is further extended in \cite{YLV14} to allow for adaptive degree distributions at the relays  based on feedback of error rate from the destination.

In this paper we focus on improving the design of DLT codes for networks with $S$ sources, $R$ relays and one destination. In particular we address the following shortcomings of the conventional DLT codes proposed in \cite{SejVukDouSenPie07acssc,SPD09,SPDI10}. We first consider lossless source-relay links, and subsequently extend to lossy source-relay-links case. The performance of DLT codes in the erasure floor region improves by allowing higher maximum degree $d_{\Gamma,\text{max}}$ of the relay-degree distribution. However, for the conventional DLT codes, the maximum relay degree is upper-bounded by the number of sources $d_{\Gamma,\text{max}} \leq S$. As a consequence these codes exhibit a high erasure floor when the number of sources is small. An additional constraint is that the sources are required to have the same information block length. As a first contribution we introduce a $D$-bit buffer at the relay for each source-relay link. The relay may now select multiple encoded bits from the same source for the relay-encoding process. Consequently, the maximum degree of the relay degree distribution is no longer limited by the number of sources, but instead by $d_{\Gamma,\text{max}} \leq DS$. Furthermore, the introduction of buffers facilitates the use of different information block sizes across sources. As a second contribution the performance of the buffer-based DLT codes is evaluated through density evolution, which is subsequently the foundation of an asymptotic analytical framework for optimization of the relay-degree distribution. In a third contribution we further enable UEP by integrating weighted-selection and expanding-window strategies into the optimization framework. For the case of lossy source-relay links a problem with the conventional DLT codes is that a source may be selected for encoding at the relay; however, the particular source bit may be erased, and thus the effective degree diminishes. As a minor contribution we introduce a one-bit buffer for each source-relay link to alleviate this problem. However, the $D$-bit buffer to increase the maximum degree of the relay-degree distribution in our proposed buffer-based DLT codes will inherently alleviate this problem. Numerical examples show that there is virtually no loss in erasure rate performance between transmission over lossy and lossless source-relay links for an optimized buffer-based DLT code. Additional delays, however, may occur. The design framework and our contributions are demonstrated by a number of illustrative examples, showing the improvements obtained by the proposed buffer-based DLT codes.

The organization of the paper is as follows. In Section \ref{Sec:System} we provide an overview of LT codes and their extension to enable UEP. Furthermore, the system model for a network with $S$ sources, $R$ relays and one destination is defined. We consider the analysis and design of buffer-based DLT codes with modified encoding for a network with $S$ sources, a single relay and single destination in Section \ref{Sec:Lossless}, for the lossless source-relay links. The extensions to UEP and $R$ relays are also discussed. The extensions to lossy source-relay links are detailed in Section \ref{Sec:UEP_lossy_single}, and numerical results, demonstrating the improvements of the proposed DLT codes, are presented in Section \ref{Sec:Numerical}. Conclusions are drawn in Section \ref{Sec:Conclusion}.


\section{Preliminaries and System Model}
\label{Sec:System}
\subsection{LT Codes}
In an LT code a vector $\mathbf{u}=[u_1,u_2,\ldots,u_K]$ of $K$ information bits (also termed variable nodes) is encoded into a potentially unlimited number of coded bits $\mathbf{c}=[c_1,c_2,\ldots]$ (also termed check nodes). As such an LT code can be regarded as a series of rate-compatible instances of irregular LDGM codes \cite{LDGM2003}, described by a generator matrix that has a new row added for each new coded bit produced and transmitted. The $t$-th coded bit $c_t$ is generated by first randomly selecting a degree $j$ from a predetermined check-node degree distribution $\Omega(x)=\sum_{j=1}^{d_{\Omega, \text{max}}}\Omega_j x^j$, where $\Omega_j$ is the probability of a degree $j$ check node being chosen and $d_{\Omega, \text{max}} \leq K$ denotes the maximum check-node degree; subsequently $j$ information bits are selected uniformly-at-random from the $K$ information bits and added modulo-2. The resulting coded bit $c_t$ is then transmitted over the channel.

For transmission over a BEC, some of the $N$ transmitted bits (corresponding to an instantaneous rate of $R=K/N$) are erased and hence decoding is realized only on the $\widehat{N} \leq N$ correctly received coded bits. The columns in the generator matrix corresponding to the correctly received coded bits describe the relationship at the decoder between the received coded bits and the information bits. This relationship can also be represented by a Tanner graph with $K$ variable nodes (corresponding to the $K$ information bits) and $\widehat{N}$ check nodes (corresponding to the received coded bits), where potentially $\widehat{N}\rightarrow \infty$. Each check node is representing a column in the generator matrix, and is therefore connected to the $j$ variable nodes randomly selected to generate the coded bit. The resulting graph is termed the decoding graph. Once a sufficient number of coded bits are received, decoding is performed using iterative message passing (based on the belief propagation algorithm) over the decoding graph to recover the information bits. As soon as the decoder has recovered all information bits an acknowledgment is sent to the transmitter to halt the current session of transmission.

As the information bits are selected uniformly-at-random by the check nodes, and since the BEC erases transmitted bits uniformly-at-random, the check-node degree distribution $\Omega(x)$ remains the same for the encoding and decoding graphs. This is not the case for the variable-node degree distributions. As only the decoding graph is relevant for our analysis, we just provide the variable-node degree distribution at the decoder. Exchanging $\widehat{N}$ with $N$ provides the corresponding variable-node degree distribution at the transmitter.

The corresponding variable-node degrees are described by a binomial distribution. As $K$ grows large the binomial distribution can be accurately approximated by a Poisson distribution \cite{Sho06}. Consequently, the variable-node degree distributions can be determined for a particular instance of $\widehat{N}$, and subsequently approximated by
\begin{equation}\label{eqn:vardeg}
\Lambda(x) = \sum_{i=1}^{d_{\Lambda, \text{max}}} \Lambda_i x^i \approx e^{\mu(x-1)}.
\end{equation}
Here $d_{\Lambda, \text{max}} \leq \widehat{N}$ is the maximum variable-node degree at the decoder, and $\Lambda_{i}$ is the probability that a particular variable node is of degree $i$, determined as
\begin{equation}\label{eqn:vardeg1}
\Lambda_i= \dbinom{\widehat{N}}{i} \left(\frac{\mu}{\widehat{N}}\right)^i\left(1- \left(\frac{\mu}{\widehat{N}}\right)\right)^{\widehat{N}-i} \approx\frac{e^{-\mu }\mu^{i}}{i!},
\end{equation}
where
\begin{equation}\label{eqn:vardeg2}
\sum_{i=1}^{d_{\Lambda, \text{max}}}\Lambda_{i}=1\hspace{5mm} \text{and}\hspace{5mm} 0\leq\nonumber \Lambda_{i}\leq1.
\end{equation}
The involved Poisson parameter $\mu$ is determined as
\begin{equation}\label{eqn:vardeg3}
\mu=\Omega_{\text{avg}}\frac{\widehat{N}}{K},
\end{equation}
where $K$ and $\widehat{N}$ are assumed to be asymptotically large with a fixed ratio $\widehat{N}/K$, and $\Omega_{\text{avg}}$ is the average check-node degree.

The above degree distributions are defined from a node-perspective. For asymptotic performance analysis of graph-based codes, edge-perspective degree distributions are more convenient. For an LT code we denote the edge-degree distribution of the variable nodes and check nodes by $\lambda(x)$ and $\omega(x)$, respectively. Here $\lambda_i$ is the probability that a variable-node edge emanates from a degree-$i$ variable node and $\omega_j$ is the probability that a check-node edge emanates from a degree-$j$ check node. The relationship of the two respective degree distributions is given by,
\begin{eqnarray}\label{eqn:vnedgedeg}
\omega(x)=\frac{\Omega'(x)}{\Omega'(1)}=\sum_{j=1}^{d_{\Omega, \text{max}}}\omega_j x^{j-1},\\
\lambda(x)=\frac{\Lambda'(x)}{\Lambda'(1)}=\sum_{i=1}^{d_{\Lambda, \text{max}}}\lambda_i x^{i-1}\approx e^{\mu(x-1)},
\end{eqnarray}
where $f'(x)$ is the derivative of $f(x)$ with respect to $x$.


\subsection{LT Codes with UEP}
\label{Sec:UEP}
Two strategies have been suggested for constructing LT codes with unequal error protection. In both cases, the information bits are divided into a number of importance classes, $\eta_1, \eta_2,..., \eta_I$, where each class contains the information bits of a certain level of importance (e.g., reliability). The number of information bits in class $\eta_i$ is denoted by $\kappa_i$ and clearly $\sum_{i=1}^{I} \kappa_i = K$.

The approach suggested in \cite{RVF07} is based on a biased (non-uniform) selection of bits from the different importance classes for the encoding process, typically referred to as weighted-selection. The bits in class $\eta_i$ has a higher probability of being selected than bits in class $\eta_j$
to ensure stronger protection for the bits in class $\eta_i$. In contrast, the approach proposed in \cite{SVDSP09} is based on a uniform selection of bits from a set of progressively expanding subsets of information bits. Here, the first subset, $\varpi_1$, contains the bits in class $\eta_1$, while subset $\varpi_i$ contains the bits from classes $\eta_1, \eta_2,...,\eta_i$. This partition determines a sequence of strictly increasing subsets of information bits, which are denoted as windows \cite{SVDSP09}. The division into importance classes is described by the importance-distribution $\Pi(x) = \sum_{i=1}^{I} \Pi_i x^i$, where $\Pi_i = \frac{\kappa_i}{K}$. As part of the encoding process, a coded bit is randomly assigned to a window $\varpi_i$ with probability $\theta_{i}$, described by the window-assignment distribution $\theta(x) = \sum_{i=1}^{I} \theta_i x^i$. Also a particular check-node degree distribution, $\Omega^{(i)}(x) = \sum_{j=1}^{\kappa_j} \Omega^{(i)}_j x^j$, is assigned to each window. 
The encoding progresses as follows. First the window-assignment distribution is sampled to determine the window. Then the degree distribution $\Omega^{(i)}(x)$ is sampled and $d$ information bits are subsequently selected uniformly-at-random within the window $\varpi_i$, corresponding to the sampled degree. The overall UEP coding scheme is termed expanding-window LT codes (EWLT) and can be seen as a generalization of conventional LT codes \cite{LT} and UEP-LT codes \cite{RVF07}.


\subsection{System Model}
Wireless erasure networks, as thoroughly investigated in \cite{DGPHE06}, are of particular interest for two reasons: Firstly they are a meaningful abstractions of real-world systems; and secondly they form a class of networks which, in general, allows for mathematically tractable problem formulations. The erasure channel models the fundamental detrimental effects of deep fade events experienced over a wireless channel. Likewise, in the higher layers of the communication protocol the link model is typically a packet erasure channel, where a data packet is either received perfectly or not received at all due to buffer overflows, excessive delays, or erroneous packets caused by the lower layers. The symbol cardinality for an LT code can be arbitrary, from binary symbols to general $2^{\ell}$-ary symbols. In the general case a $2^{\ell}$-ary symbol can be considered as a data packet with $\ell$ bits. In the corresponding encoder/decoder processing bit-wise module-2 processing is applied to each symbol/packet. For notational simplicity here we have assumed binary symbol; however the analysis is also valid for packets having arbitrary size.

Our objectives are to investigate the performance of a relay network having $S \geq 2$ sources and $R \geq 1$ relays using DLT codes, and subsequently develop a design framework for optimization. Note that the use of multiple relays in the network provides resilience against relay node failures, e.g. due to power loss. In such situations, active relays in the network forward information to the destination irrespective of the status of other relays. The system model is defined as follows. The transmission through the network is scheduled into transmission rounds. Each round consists of two phases; a source transmission phase and a relay transmission phase. In each phase, the allocated time interval is divided into time slots; $S$ time slots for the source transmission phase, and $R$ time slots for the relay transmission phase. Thus, each source and each relay have a time slot for transmission in each phase, respectively.

\begin{figure}
  \begin{center}
    \includegraphics[width=75mm]{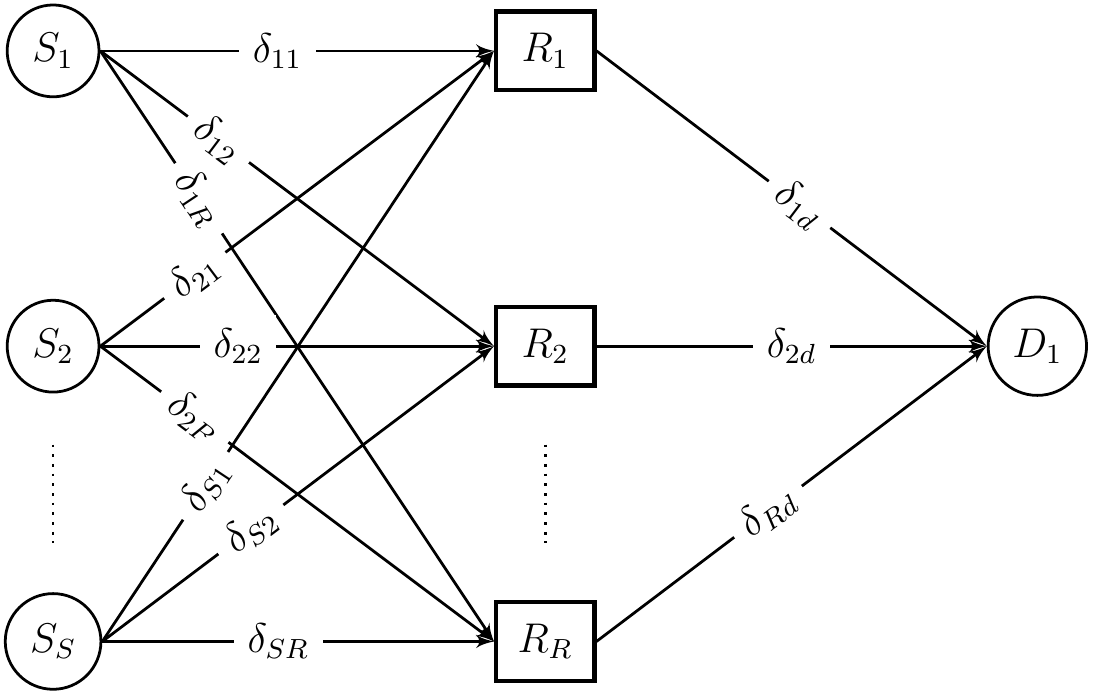}
    \caption{System Model for $S$ Sources and $R$ Relays Network.}\label{System_Model}
    \end{center}
\end{figure}

Each source has a total of $K_i$, $i=1,2,\ldots,S$ information bits to transmit to a common destination through a set of relays, as shown in Fig.~\ref{System_Model}. Note that $K_i$ may not be equal to $K_j$ for $i\neq j$. Each source further encodes its information bits into coded bits using a common check-node degree distribution\footnote{Different degree distributions can be used at the sources. However, for notational ease, and simplified density evolution expressions, we assume a common degree distribution across all sources. Throughout the paper, unless explicitly stated otherwise, we use the source check-node degree distribution detailed in \cite{SPD09} for encoding the information bits of sources.} $\Omega(x)$. In round $n$ the coded bit transmitted from source $i$ to the relays is denoted by\footnote{The superscript $n$ is used throughout for indicating the corresponding transmission round for the parameter concerned.} $c_{i}^{n}$. We assume that the link between source $i$ and relay $j$ is a BEC with erasure probability $\delta_{ij}$, where $i=1,2,\ldots,S$ and $j=1,2,\ldots,R$.

Relay $j$ receives the coded bits transmitted from all the sources during the source transmission phase as a set $[\beta_{1j} c^n_{1}, \beta_{2j} c^n_2, \dots, \beta_{ij} c^n_i, \dots, \beta_{Sj} c^n_{S}]$, where $\beta_{ij}$ is a Bernoulli random variable representing the BEC link connecting source $i$ and relay $j$. Here, $\beta_{ij}=1$ with probability $(1-\delta_{ij})$ and $\beta_{ij}=0$
with probability $\delta_{ij}$ indicating that $c_{i}^{n}$ is
received/erased at relay $j$. In round $n$ relay $j$ generates a coded bit $z_{j}^{n}$ by combining a random number of the coded bits received from the sources.
The number of coded bits to combine is determined by sampling the relay-degree distribution, represented by $\Gamma_j(x)=\sum_{d=1}^{d_{\Gamma_{j}, \text{max}}} \Gamma_{j,d} x^d$. Here $\Gamma_{j,d}$ is the probability that a newly generated relay-coded bit combines the source-coded bits from $d$ sources, where $d_{\Gamma_{j}, \text{max}}$ is the maximum number of sources to be selected.
Subsequently the relay-coded bits are transmitted to the destination in the relay transmission phase. The link between relay $j$ and the destination is also a BEC with erasure probability $\delta_{jd}$. In the literature on rateless codes, the asymptotic analysis is predominantly conducted in terms of reception overhead $\varepsilon_r=\widehat{N}/K$, e.g., \cite{SPD09,LiauKimYousefi13TCOM}. Likewise the resulting erasure rate performance is commonly measured against the reception overhead. In our analysis we follow the conventional approach, expressing the asymptotic performance in terms of the reception overhead; however we measure the resulting erasure rate performance of our proposed DLT codes as a function of the overall transmission overhead $\varepsilon=N/K$ where $K=\sum_{i=1}^S K_i$ and $N$ is the number of transmitted coded bits from each source at the instance of decoding. Due to the averaged relationship between transmission and reception overhead, the effect is merely a horizontal shift.


\section{Lossless Source-Relay Links}
\label{Sec:Lossless}
For ease of understanding, we first consider EEP in a network with $S$ sources and a single relay in Subsection \ref{Subsec:EEP_lossless_single}. To simplify the initial system further we assume lossless source-relay links. We subsequently generalize the lossless case to an $S$-sources single relay system providing UEP across all the sources, and to the similar  case of an $S$-sources $R$-relay network in Subsections \ref{Sec:UEP_lossless_single} and \ref{Subsec:UEP_lossless_R}, respectively. The case of lossy source-relay links is considered in Section \ref{Sec:UEP_lossy_single}.


\subsection{EEP for a Single-Relay Network}
\label{Subsec:EEP_lossless_single}
Here we consider the design of DLT codes for lossless source-relay links ($\delta_{ij}=0$) in a network with $S$ sources and a single relay. With a single relay the notation is simplified since $\beta_i=\beta_{i1}=1$ for $i=1,2,\ldots,S$; the relay-destination link erasure probability is $\delta=\delta_{1d}$, the relay-degree distribution is denoted by $\Gamma(x)=\Gamma_1(x)$; and the coded bit transmitted from the relay to the destination during round $n$ is denoted by $z^{n}$. In this subsection we only consider the case of EEP across the sources.

\subsubsection{Conventional DLT Codes}
The DLT codes proposed in \cite{SPD09}, which we denote as the conventional DLT codes, are designed for a multi-source single relay network. At any given transmission round, each source transmits an encoded bit to the relay in the corresponding designated source time slot. 
The relay then uniformly-at-random combines $d$ encoded bits received from different sources, where the number of sources $d$ to be selected is sampled from the relay-degree distribution $\Gamma(x)$. In round $n$ the encoded bit at the relay is determined as
\begin{equation}\label{}
z^n=\overset{d_{\Gamma, \text{max}}}{\underset{i=1}{\bigoplus}} \zeta_i c^n_i,
\end{equation}
where $\bigoplus$ denotes modulo-2 summation. Here, $\zeta_i=1$ if source $i$ is selected for the combining process and $\zeta_i=0$ otherwise. The encoded bit $z^n$ is subsequently transmitted from the relay to the destination over a BEC with channel erasure probability $\delta$.  The transmission continues until the destination has successfully received all information bits from all sources. Decoding is performed over a graph containing variable nodes from all sources and as many correctly received check nodes from the relay at the instance of decoding.

It has been shown in \cite{SPD09} that the performance of the DLT codes in the erasure floor region improves as the maximum degree of the relay-degree distribution grows. However, the maximum degree of the relay-degree distribution is upper-bounded by the number of sources $d_{\Gamma,\text{max}}\leq S$. A recognized drawback is therefore a high erasure floor when the number of sources is small. An additional constraint is that the sources are required to have the same information block length. As the relay selects the source bits for encoding with equal probability, the conventional DLT codes cannot provide EEP performance for sources with different information block lengths without modifications to the relay-degree distribution.

\subsubsection{Proposed DLT Codes}
To address the shortcomings noted above we introduce a set of buffers $B_i$, $i=1,2,..., S$, of size $D$ bits; one for each source-relay link. The encoded bits stored in buffer $B_i$ are denoted by $[b^1_i,b^2_i,\ldots,b^D_i]$ where $b^m_i$ is the $m$-th bit of buffer $B_i$. All the buffers operate in a first-in-first-out mode where a right-shift action is performed with every new entry into the buffer. At the completion of source round $n$, the relay has received new encoded bits from all active sources and the corresponding buffers are updated. The newly received coded bit $c_{i}^{n}$ is right-shifted into the respective buffer $B_i$ at position $b^1_i$. As a result all the previously stored bits in buffer $B_i$ are right-shifted causing the encoded bit $c_{i}^{n-D}$ previously stored in the right-most buffer position $b^D_i$ to be discarded. Due to the buffers, the relay may now select multiple encoded bits from the same source for the relay-encoding process of relay-coded bit. Consequently, the maximum degree $d_{\Gamma, \text{max}}$ of $\Gamma(x)$ is no longer limited by the number of sources in the network. Furthermore, the introduction of buffers facilitates the use of different information block sizes across sources; however, an initial delay of $D$ transmission rounds is required to ensure that all buffers are filled, after which the relay can transmit properly encoded bits in each relay transmission phase without any further delay. This delay may lead to a slightly higher transmission overhead for a given erasure rate as compared to the existing DLT coding schemes. Typically $D \ll N$ so the initial delay before the relay starts transmission is quite small and subsequently the additional transmission overhead is negligible.

When all buffers are updated, the relay samples the relay-degree distribution $\Gamma(x)$ to determine the number of buffered bits required for the encoding process of a new relay-coded bit. We define the source-selection probability at the relay as $q(x)=\sum_{i}^{S}q_i x^i$, where $q_i$ is the probability that a bit from buffer $B_i$ is selected for the current encoding process at the relay. We also define the fraction of variable nodes of source $i$ as $\alpha_i=K_i/K$, where it follows that $\sum_{i=1}^{S} \alpha_i=1$. Finally we define the bias factor $w_i=q_i/\alpha_i$ for $i=1,2,..., S$, which determine the distribution of UEP across all sources. If $w_1=w_2,\ldots,w_S=1$ the bits from all source are equally protected. In contrast, if $w_i > 1$ the bits from source $i$ are better protected than the EEP level, and if $w_i < 1$ the bits from source $i$ are worse protected than the EEP level. Since $q_i=1/S$ for the conventional DLT codes it is clearly not possible to obtain $w_1=w_2,\ldots,w_S=1$, corresponding to EEP performance, for sources with different information block sizes.

\begin{figure}
  \begin{center}
        \includegraphics[width=65mm]{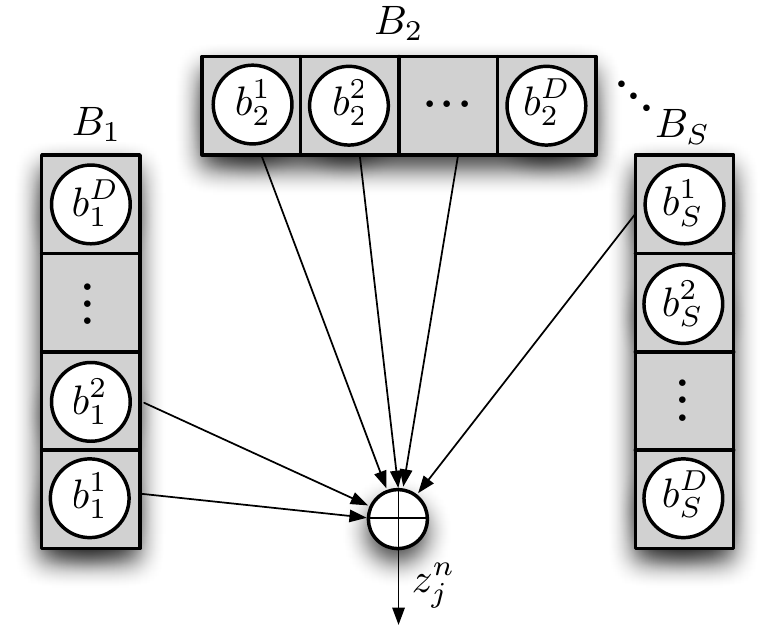}
    \caption{Combining process at relays.}\label{Relay_Combining}
  \end{center}
\end{figure}

We first consider the design of our proposed buffer-based DLT codes for the case of EEP performance. In order to meet the above constraints on $w_i$, the source selection probability $q(x)$ can be adjusted accordingly to accommodate different information block sizes of the sources. Following the sampling of the relay-degree distribution $d$ source bits are selected for encoding from among all the source bits currently in the buffers. The source selection probability $q(x)$ is then sampled $d$ times to determine how many bits $d_i \leq d$ to be selected from buffer $B_i$ for $i=1,2,\ldots,S$, where $d=\sum_{i=1}^S d_i$. If $d_i$ bits are selected from buffer $B_i$, buffer bits $[b^1_i,b^2_i,\ldots,b^{d_i}_i]$ are selected for encoding to give high priority to the newly received coded bits from the sources. Finally, all the selected buffer bits from different buffers are added modulo-2 as
\begin{equation}\label{Rbit}
z^n=\overset{S}{\underset{i=1}{\bigoplus}} \psi_i\left(\overset{d_i}{\underset{m=1}{\bigoplus}}b^{m}_{i}\right),
\end{equation}
and transmitted over the relay-destination link. Here $\psi_i=1$ if buffer $B_i$ is selected and $\psi_i=0$ otherwise. The corresponding combining process at the relay is shown in Fig.~\ref{Relay_Combining} and detailed in Algorithm~\ref{alg1}. The combining process at the relay is demonstrated by the following example.

\begin{example}
Consider a single-relay network with $S=4$, $D=8$. At round $n$, $\Gamma(x)$ is sampled with outcome $d=7$, and $q(x)$ is sampled 7 times with accumulated outcome $d_1=1$, $d_2=3$, $d_3=0$ and $d_4=3$. All the buffers are subsequently updated with right-shift operation such that $b^1_i=c_{i}^{n}$ for buffer $B_i$ where $i=1,2,\ldots,S$. The coded bit $c_{i}^{n-8}$ which was previously stored at $b^8_i$ is removed from the buffer $B_i$. Then $z^n$ is calculated from \eqref{Rbit} as $z^n=b^1_1\oplus b^1_2 \oplus b^2_2 \oplus b^3_2 \oplus b^1_4 \oplus b^2_4 \oplus b^3_4$. Note that $d_{\Gamma, \text{max}}\geq D > S$, which is clearly not possible with conventional DLT codes.
\end{example}

\begin{algorithm}[t]                                    
\caption{: Proposed Combining Scheme at Relay}          
\label{alg1}                                            
\begin{algorithmic}
\STATE {\bf Initialization:} Wait $D$ transmission rounds to allow all buffers to be filled.
\WHILE{unless relay receives an acknowledgment from the destination to halt transmission}
\STATE {\bf Step 1: } After completion of source transmission phase in round $n$, update all buffers with right-shift operation such that $b^1_i=c_{i}^{n}, b^2_i=c_{i}^{n-1},\ldots, b^D_{i}=c_{i}^{n-D+1}$ for $i=1,2,\ldots,S$. The previously stored buffer bit $b^D_i$ in buffer $B_i$ is discarded for all $i$.
\STATE {\bf Step 2: } Sample $\Gamma(x)$ to determine the number of buffer bits $d$ to be combined to form the relay-coded bit $z^n$.
\STATE {\bf Step 3: } Sample $q(x)$ to determine the buffers and number of bits from each buffer i.e. $d_i$ for $i=1,2,\ldots,S$.
\STATE {\bf Step 4: } Calculate $z^n=\overset{S}{\underset{i=1}{\bigoplus}} \psi_i\left(\overset{d_i}{\underset{m=1}{\bigoplus}}b^{m}_{i}\right)$. 
\STATE {\bf Step 5: } Update the generator matrix at the relay.
\STATE {\bf Step 6: } Transmit $z^n$ over relay-destination link.
\ENDWHILE
\end{algorithmic}
\end{algorithm}

Once all buffers are filled, the relay transmits continuously in every transmission round. A more serious problem is that the relay-encoded bit $z^n$ may be connected to multiple coded bits from the same source. Since the encoding process at the relay is based on modulo-2 summation, some information bits may in this case be excluded from a newly generated relay-coded bit. Hence, the probability that a certain number of information bits remain unconnected may be higher for the proposed codes as compared to the conventional DLT codes. We demonstrate this problem with the following example.

\begin{example}
Consider a single-relay network with $S=2$ and $D=4$. At round $n$, $\Gamma(x)$ is sampled with outcome $d=2$, and $q(x)$ is sampled twice with outcome $d_1=2$ and $d_2=0$. Further, let  $c_{1}^{n}=u^{10}_1$ and $c_{1}^{n-1}= u^{10}_1\oplus u^{15}_1$. In this case $z^n$ is calculated as
\begin{equation*}
z^n = b^1_1\oplus b^2_1 = c_{1}^{n}\oplus c_{1}^{n-1} = u^{10}_1\oplus u^{10}_1 \oplus u^{15}_1= u^{15}_1.
\end{equation*}
\end{example}

\begin{figure}
  \begin{center}
    \includegraphics[width=70mm]{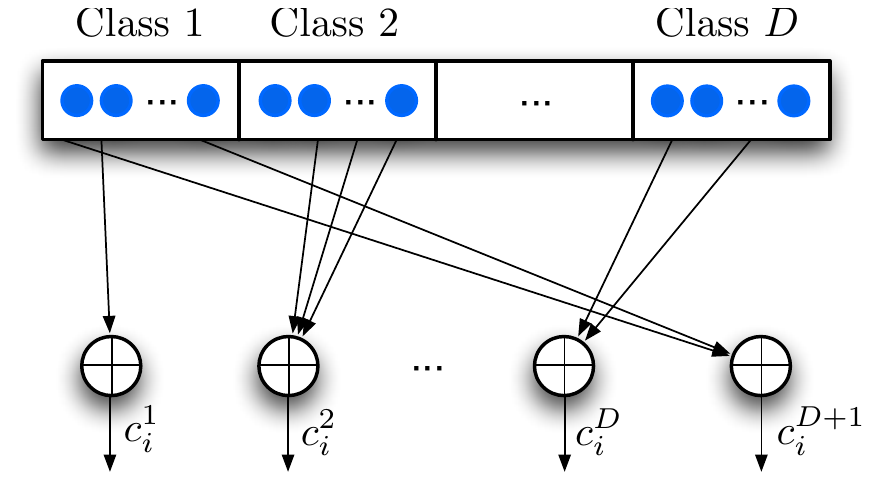}
    \caption{Encoding at source $i$.}\label{Source_Encoding}
  \end{center}
\end{figure}

In order to circumvent this problem, the encoding process is modified as follows. The information bits of each source is divided into $D$ classes, each having $\xi^\ell_i$ information bits such that $K_i= \xi^\ell_i D$ for $i=1,2,\ldots,S$ and $\ell=1,2,\ldots,D$. The encoding process for source $i$ at transmission round $n$ is then restricted to selecting only from the $\xi^\ell_i$ information bits in class $v_\ell = \text{mod} (n-1, D)+ 1$. It follows that the information bits of source $i$, $i=1,2,...,S$, involved in the source encoding process of $D$ consecutive source-encoded bits, belong to different classes, $v_1, v_2,..., v_\ell,...,v_D$ at the source. Thus, if multiple coded bits are selected from the same source at the relay, their corresponding information bits belong to different classes. Having $D$ classes, we are ensured that self-cancellation cannot occur. The sequential encoding process is illustrated in Fig.~\ref{Source_Encoding}. To further clarify our proposed scheme, the modified encoding process at the sources is detailed in Algorithm~\ref{alg2}.

\begin{algorithm}[t]                                               
\caption{: Modified Encoding Scheme at Sources}                    
\label{alg2}                                                       
\begin{algorithmic}
\STATE {\bf Initialization:} At source $i$ for $i=1,2,\ldots,S$, divide all the information bits into $D$ classes, each having $\xi^\ell_i$ information bits such that $K_i=\xi^\ell_i D$ with $\ell=1,2,\ldots,D$ corresponding to each source.
\WHILE{unless an acknowledgment is received from the destination}
\STATE {\bf Step 1: } To generate a coded bit $c_{i}^{n}$ in source transmission round $n $, select a degree $h_i$ from $\Omega(x)$ for $i=1,2,\ldots,S$.
\STATE {\bf Step 2: } Determine the class $v_\ell=\text{mod}(n-1,D)+1$ from which information bits are to be selected for $c_{i}^{n}$ for all $i$.
\STATE {\bf Step 3: } At each source select $h_i$ information bits uniformly-at-random from their respective class $v_\ell$.
\STATE {\bf Step 4: } Calculate $c_{i}^{n}$ by modulo-2 summation of the selected $h_i$ information bits for $i=1,2,\ldots,S$ and transmit to the relay.
\ENDWHILE

\end{algorithmic}
\end{algorithm}

\subsubsection{Optimal Relay-Degree Distribution for EEP Performance}
The information bits from each class are selected uniformly-at-random during the encoding process, and thus the variable-node degrees in each class are binomial distributed. It follows that the variable-node degree distribution in each class can be approximated asymptotically by a Poisson distribution \cite{Sho06}. As a result, the variable node degrees are Poisson distributed at each source. Similarly, since the selection of coded bits from the sources for the relay encoding process is performed uniformly-at-random, a Poisson distribution can also approximate the variable-node degree distribution at the destination \cite{SPD09}. The probability that a variable node is not recovered at the destination after $l$ decoding iterations is accordingly given by
\begin{eqnarray}\label{eqn:DEEEP}
P_{0}&=& 1, \nonumber\\
P_{l} & \approx& \exp[-\bar{\mu}\omega(1-P_{l-1}) \gamma(\Omega(1-P_{l-1}))],
\end{eqnarray}
where $\bar{\mu}=\Gamma'(1)\Omega'(1)\varepsilon_r$ is the average variable-node degree at the decoder and $\omega(x)$, $\gamma(x)$ are the edge-perspective degree distributions corresponding to $\Omega(x)$, $\Gamma(x)$, respectively. The reception overhead can be expressed in terms of the edge-perspective relay-degree distribution as $\varepsilon_r=\frac{\bar{\mu}}{\Omega'(1)}\sum_{j=1}^{d_{\Gamma, \text{max}}}\gamma_j/j$. Thus the  expression in \eqref{eqn:DEEEP} can be transformed into a linear program when $\Omega(x)$ is known. The corresponding linear program is formulated as follows
\[ \begin{array}{rll} \label{eqn:rholinprog}
&\text{LP1}\\
&\mbox{minimize} & \frac{\bar{\mu}}{\Omega'(1)}\sum_{j=1}^{d_{\Gamma, \text{max}}} \gamma_j/j \\
&\mbox{subject to } &\gamma_j\geq 0, \\
&&\sum_{j=1}^{d_{\Gamma, \text{max}}} \gamma_j=1,\\
&&\sum_{j=1}^{d_{\Gamma, \text{max}}}\gamma_{j}\Omega(x_i)^{j-1}\geq-\frac{-\ln(1-x_i)}{\bar{\mu}\omega(x_i)},\\
\end{array}
\]
note that $i\in 1,2,\ldots,m$ and $0=x_1<x_2<\ldots<x_m=1-\epsilon$ are equidistant points on $[0, 1-\epsilon]$, where $\epsilon$ is the desired erasure rate. We use the linear program LP1 to obtain the optimized edge-perspective relay-degree distribution $\gamma(x)$ for our proposed DLT code in terms of minimum reception overhead (and consequently also in terms of minimum transmission overhead). This degree distribution is then converted to the node-perspective relay-degree distribution $\Gamma(x)$ for the combining process at the relay.  For our proposed scheme $d_{\Gamma, \text{max}}>S$ is possible as compared to the conventional DLT coding scheme where $d_{\Gamma, \text{max}}\leq S$.


\subsection{UEP for a Single-Relay Network}
\label{Sec:UEP_lossless_single}
To provide UEP performance between the sources with our proposed DLT codes, we consider the two main approaches previously discussed in Section \ref{Sec:UEP}; namely weighted selection \cite{RVF07} and expanding windows with uniform selection \cite{SVDSP09}. Unequal protection of information bits within a source is easily done by modifying the common check-node degree distribution $\Omega(x)$ accordingly. We therefore focus on unequal protection between sources provided through the relay encoding process.

\subsubsection{Optimized UEP Using Weighted-Selection DLT Codes}
It is a challenging task to provide UEP performance for DLT codes in a multi-source relay network. However the relay combining process proposed in \eqref{Rbit}, and detailed in Algorithm~\ref{alg1}, can be modified to obtain UEP performance across sources with our proposed DLT codes. To realize UEP, we introduce a biased source selection process at the relay such that $w_i>w_j$ if source $i$ is to be better protected than  source $j$. This approach is able to achieve UEP performance irrespective of the number of sources, and the sizes of information blocks at the sources. Here, the set of parameters $w_i$ for $i=1,2,...,S$ are selected to meet the constraints defined by the importance classification. The probability $P^{W}_{l,i}$ that a variable node in source $i$ is not recovered after $l$ decoding iterations for $i=1,2,\ldots,S$ is given as
\begin{eqnarray}\label{DEUEP}
P^{W}_{0,i}&=&1, \\
P^{W}_{l,i} &\approx& \exp[(-\bar{\mu_i}\omega(1-P_{l-1,i})\gamma(\sum_{m=1}^Sq_m\Omega(1- P_{l-1,m})))], \nonumber
\end{eqnarray}
where $\bar{\mu_i}=w_{i}\bar{\mu}$ is the average variable-node degree of source $i$. Furthermore, it is straightforward to derive a lower bound for each source of our proposed DLT codes based on ML decoding, following a similar approach as detailed for LT codes in \cite{RVF07}.

\begin{figure}
  \begin{center}
        \includegraphics[width=65mm]{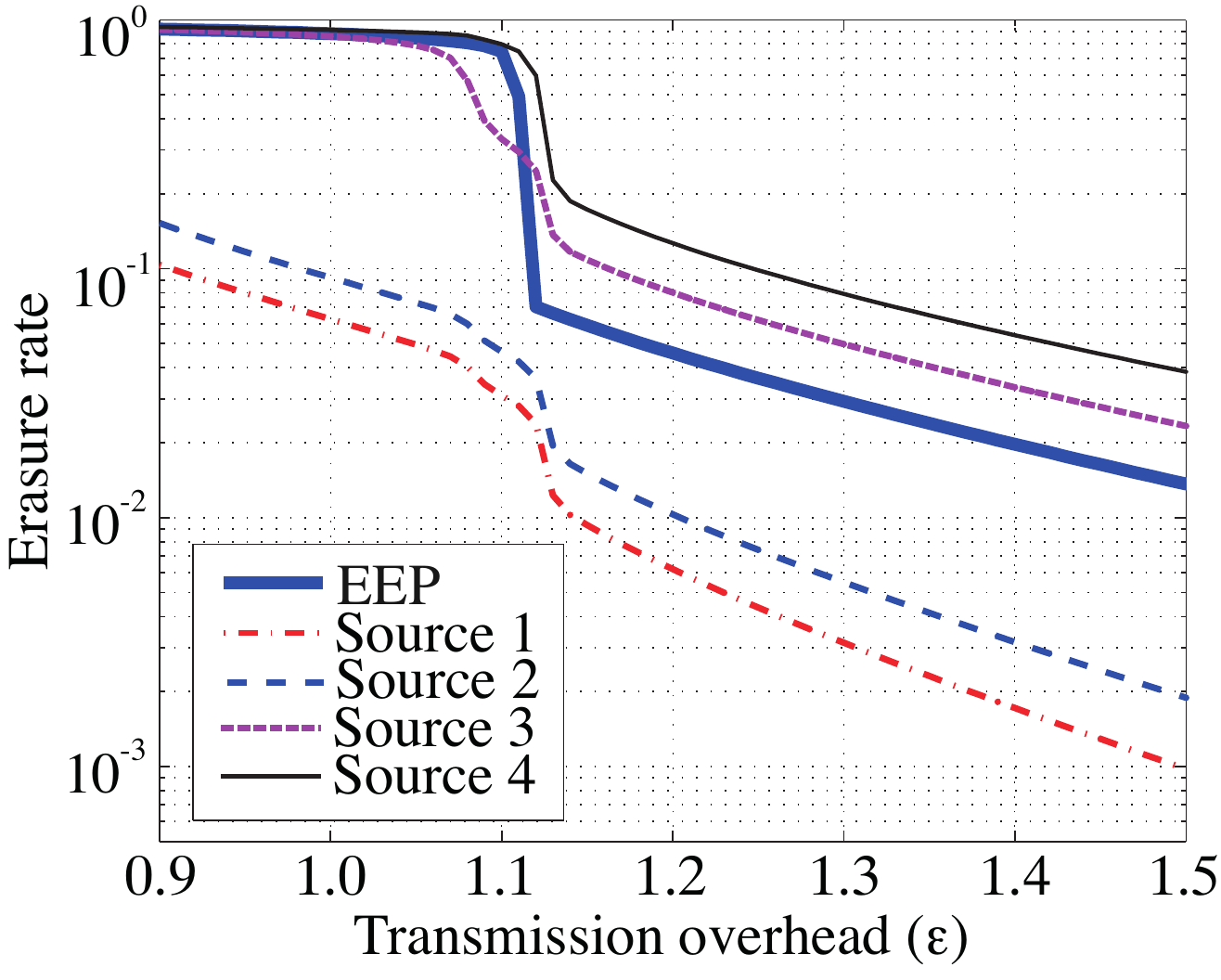}
    \caption{Asymptotic UEP performance of DLT codes for a four-sources, single-relay scenario with relay-degree distribution optimized for EEP.}\label{Th_UEP_S4_R1_Lossless_1}
  \end{center}
\end{figure}
As an example we consider a four-sources scenario with the assumption of perfect source-relay links, and the relay-destination link being a BEC with erasure probability $\delta=0.1$. We further assume that $\alpha_1=0.05$, $\alpha_2=0.20$, $\alpha_3=0.30$ and $\alpha_4=0.45$. The source-selection probability function is specified as
\begin{equation}\label{sDD}
q(x)=0.08x+0.29x^2+0.27x^3+0.36x^4.
\end{equation}
 It follows that source $i$ is better protected than source $j$ if $w_i>w_j$ for $i<j$. For the combining process at the relay we use the relay-degree distribution $\Gamma(x)$ determined by LP1 for $d_{\Gamma, \text{max}} = 4$ and given by
\begin{equation}\label{rDD}
\Gamma(x)=0.7520x+0.1685x^2+0.0455x^3+0.0340x^4.
\end{equation}
For this example  we use $D=4$ to allow multiple selection from the same source. The corresponding asymptotic UEP performance is shown in Fig.~\ref{Th_UEP_S4_R1_Lossless_1}. For comparison, we also plot the optimized EEP asymptotic performance of a four-sources scenario. Since $w_i>1$ for $i=1,2$ we observe that the first two sources exhibit better performance than the EEP performance. The remaining two sources have higher erasure rates than the EEP counterpart. It is possible to obtain a wide range of UEP alternatives across the sources, given the limits of the overall network performance. In particular, we can achieve any level of protection for a specific source $i$ by proper design of the parameter $w_i$. Hence, our proposed UEP scheme for DLT codes can easily be extended to any number of sources and any level of protection trade-offs.

The relay-degree distribution in \eqref{rDD} is optimized for EEP performance. However, taking into consideration the selection probabilities and the information bit block sizes of the sources, the density evolution expression in \eqref{DEUEP} can be reformulated into the following linear program, enabling the optimization of the relay-degree distribution for UEP performance.
\[ \begin{array}{rll} \label{eqn:rholinprog}
&\text{LP2}\\
&\mbox{minimize}& \frac{\bar{\mu}}{\Omega'(1)}\sum_{j=1}^{d_{\Gamma, \text{max}}} \gamma_j/j \\
&\mbox{subject to} &\gamma_j\geq 0, \\
&&\sum_{j=1}^{d_{\Gamma, \text{max}}} \gamma_j=1,\\
&&\sum_{j=1}^{d_{\Gamma, \text{max}}}\gamma_{j}\Omega(1\hspace{-1mm}-\hspace{-1mm}\sum_{i=1}^{S}q_iz_i^{q_i/\alpha_i})^{j-1}\hspace{-1mm}\geq -\frac{\ln(z_i)}{\bar{\mu}\omega(z_i)},\\
\end{array}
\]
\begin{figure}
  \begin{center}
    \includegraphics[width=65mm]{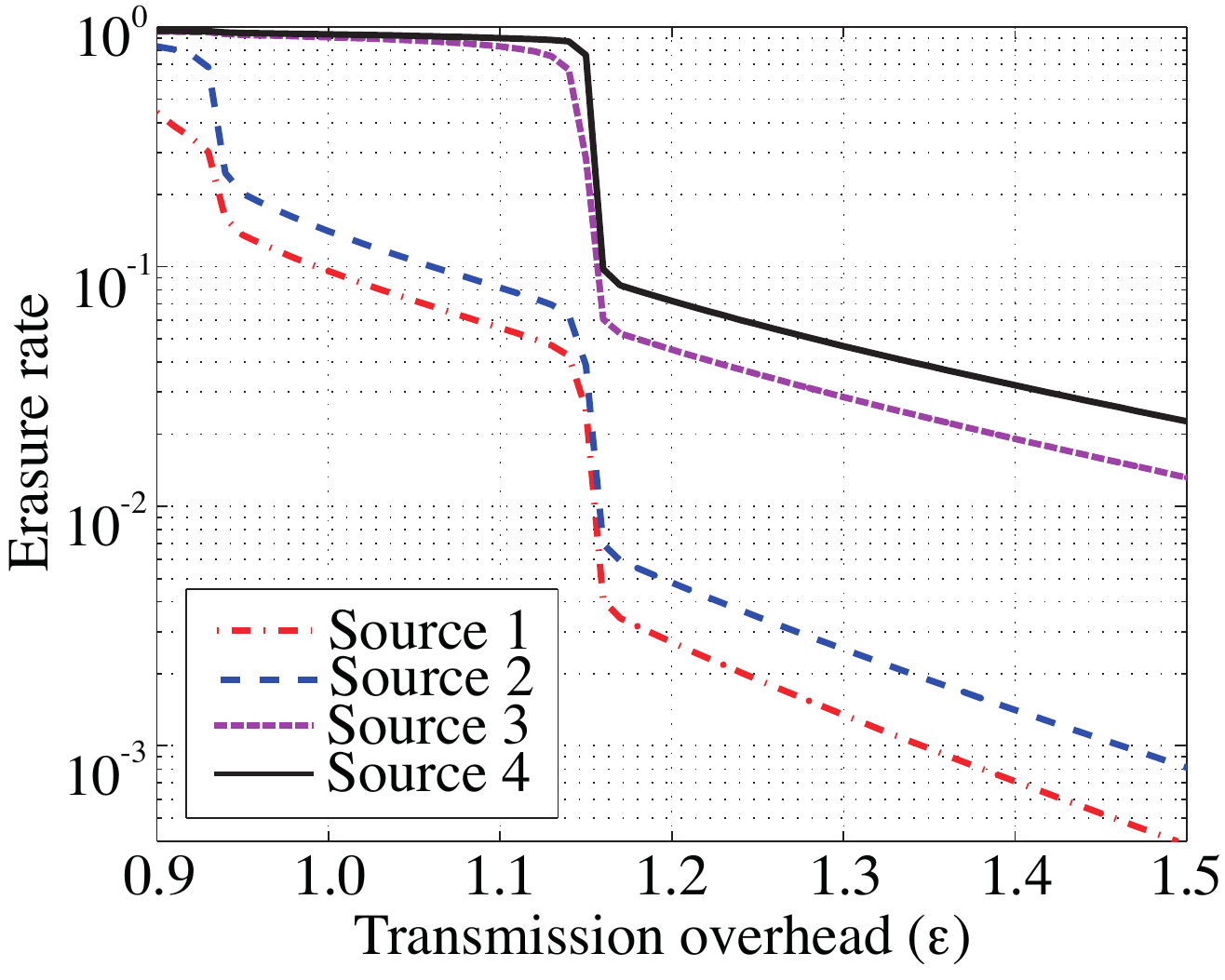}
    \caption{Asymptotic UEP performance of DLT codes for a four-sources, single-relay scenario with relay-degree distribution optimized for UEP.}\label{Th_UEP_S4_R1_Lossless_2}
  \end{center}
\end{figure}
where $1=z_1>z_2>\ldots>z_m=\epsilon$ are $m$ equidistant points on $[\epsilon,1]$. As an example we consider the optimization of a UEP-DLT code for a four-sources single-relay network. Solving the linear program LP2 for the same parameters as we solved LP1, we obtain the following relay-degree distribution,
\begin{equation}\label{rDDUEP}
\Gamma(x)=0.6021x+0.3086x^2+0.0511x^3+0.0381x^4.
\end{equation}
The corresponding asymptotic UEP performance is shown in Fig.~\ref{Th_UEP_S4_R1_Lossless_2}. Comparing  Fig.~\ref{Th_UEP_S4_R1_Lossless_1} and Fig.~\ref{Th_UEP_S4_R1_Lossless_2}, we observe that the performance of each source is improved for the case when $\Gamma(x)$ is specifically designed for unequal protection.


\subsubsection{UEP Using DEWLT Codes}
In a similar way as for weighted-selection, we can apply the expanding-window approach in \cite{SVDSP09} to provide UEP performance across the sources. The resulting codes are termed distributed EWLT (DEWLT) codes. As before, each source is assigned to an importance class. In turn, this assignment imposes an equivalent classification of the relay-coded bits, according to the expanding-window strategy. Following the approach described in Section \ref{Sec:UEP} a relay-coded bit is randomly assigned to window $\varpi_i$ with probability $\theta_{i}$, described by the window-assignment distribution $\theta(x) = \sum_{i=1}^{I} \theta_i x^i$. The corresponding relay-degree distribution assigned to window $\varpi_i$ is $\Gamma_{iW}(x)$. Sources assigned to window $\varpi_i$ is better protected than sources assigned to window $\varpi_j$ for $i < j$. To determine the asymptotic UEP performance of the proposed DEWLT codes for a single-relay network, we formulate the corresponding density evolution probability expressions $P^{EW}_{l,i}$ that a variable node in importance class $\eta_i$ is not recovered after $l$ decoding iterations as
\begin{equation}\label{DEWLT}
P^{EW}_{l,i}\hspace{-1mm}\approx\hspace{-1mm}\exp[-\varepsilon_r\hspace{-1.5 mm}\sum_{j=i}^I\frac{\theta_i}{\hspace{-1.5 mm}\sum_{t=1}^i\Pi_t}\Gamma_{jW}'(\sum_{m=1}^S q_m\Omega(1-P^{EW}_{m,l-1}))].
\end{equation}
For simplicity we consider an eight-sources single-relay network with two importance classes, where sources $1$ to $4$ are in the class of most significant bits (MSB) $\eta_1$ while sources $5$ to $8$ are in the class of least significant bits (LSB) $\eta_2$. Hence the size of the MSB class is $\kappa_1=K \sum_{i=1}^4 \alpha_i$, while the size of the LSB class is $\kappa_2= K \sum_{i=5}^8 \alpha_i$, respectively. As a result $\Pi_{1}=\sum_{i=1}^4 \alpha_i$  and $\Pi_{2}=\sum_{i=5}^8 \alpha_i$. Similarly the selection probability function for the two classes are modified accordingly as $q_{M}=\sum_{i=1}^4 q_i$  and $q_{L}=\sum_{i=5}^8 q_i$. The corresponding MSB and LSB erasure rates are determined as
\begin{eqnarray*}
P^{EW}_{l,M} \color{red}\approx \exp[-\varepsilon_r(\frac{\theta_1}{\Pi_{1}}\Gamma_{1W}'(\Omega(1-P^{EW}_{l-1,M}))\hspace{26mm}&& \\
+ \theta_2 \Gamma_{2W}'(q_{M}\Omega(1-P^{EW}_{l-1,M})\hspace{-1mm}+\hspace{-1mm}q_{L}\Omega(1-P^{EW}_{l-1,L})))],\hspace{5mm}&& \\
P^{EW}_{l,L}\hspace{-1mm} \approx \hspace{-1mm} \exp[-\varepsilon_r\theta_2\Gamma_{2W}'(q_{M}\Omega(1\hspace{-1mm}-\hspace{-1mm}P^{EW}_{l-1,M})+q_{L}\Omega(1\hspace{-1mm}-\hspace{-1mm}P^{EW}_{l-1,L}))].&&
\end{eqnarray*}

We can derive a lower bound on the erasure rate in the $i$-th class of the DEWLT code under ML decoding. This lower bound is attributed to the probability that an information bit is not connected to any relay-coded bit in the decoding graph. Recall that the number of information bits in the $j$-th window is denoted by $\kappa_{wj}=\sum_{i=1}^j\kappa_i$.
The probability that an information bit in the
$j$-th importance class is not connected to a relay-coded bit in the $j$-th window is $1-\frac{\rho_j}{\kappa_{wj}}$, where $\rho_j=\Gamma'_{jW}(1)\Omega'(1)$, provided that $j\geq i$. By averaging over the window-assignment probability $\theta(x)$, a lower bound on the probability that the ML decoder fails in recovering the variable node in importance class $\eta_i$ as a function of the overhead can be derived as
\begin{equation}\label{LBDLT}
p^{ML}_{i}(\varepsilon_r)\geq (1-\sum_{j=i}^{I} \frac{\theta_j\rho_j}{\kappa_{wj}})^{\widehat{N}}=(1-\sum_{j=i}^{I} \frac{\theta_j\rho_j}{\kappa_{wj}})^{K \varepsilon_r}.
\end{equation}

\begin{figure}
  \begin{center}
    \includegraphics[width=65mm]{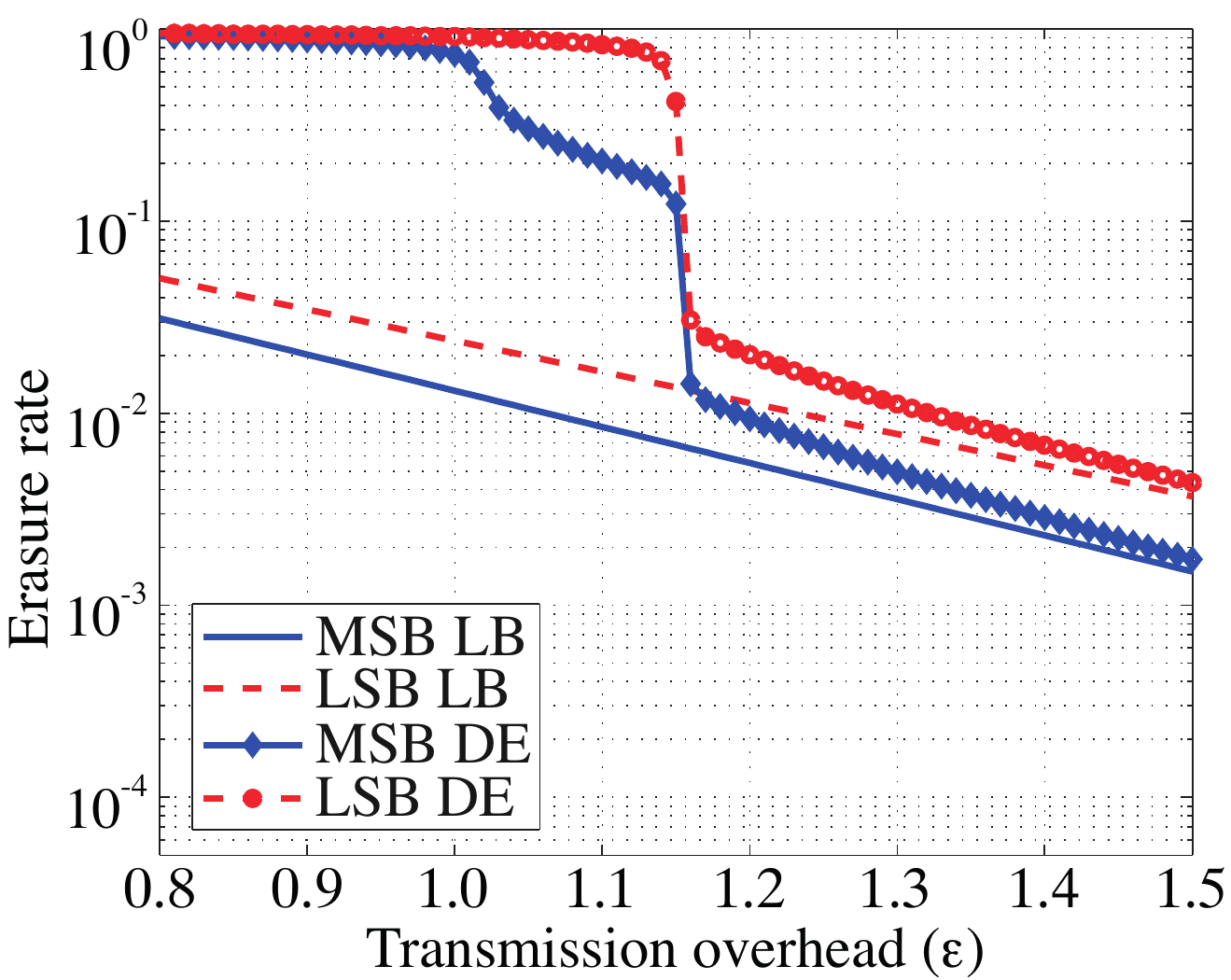}
    \caption{Lower bounds and DE of DEWLT codes for $K=24000$, $\delta=0.1$ and $\alpha_i=1/8$.}\label{LB_vs_DE}
  \end{center}
\end{figure}

The corresponding lower bounds and asymptotic performance for the two importance classes are shown in Fig.~\ref{LB_vs_DE}, where the difference in performance between the MSB and LSB classes is clear. For $\Gamma_{1W}$, the optimal degree distribution detailed in \eqref{rDD} is used, while for $\Gamma_{2W}$ optimal degree distribution is obtained from LP1 for an eight-sources scenario.

\subsection{UEP for a Generalized $R$-Relays Network}
\label{Subsec:UEP_lossless_R}
We now consider the use of our proposed UEP DLT codes in the general setting of $S$ sources transmitting to a single destination via $R$ relays as shown in Fig.~\ref{System_Model}. The weighted-selection
scheme is considered first. As before, all the sources  broadcast coded bits to all relays using a common source-degree distribution $\Omega(x)$.
In turn relay $j$ combines the source-coded bits in the corresponding buffers according to the relay-degree distribution $\Gamma_j(x)$ and forwards the relay-$j$-coded bit to the destination. As a result the probability $P^{\text{Gen},W}_{l,i}$ that a variable node of source $i$ is not recovered at the destination after $l$ decoding iteration can be determined as
\begin{equation}\label{GUEP1}
{P^{\text{Gen},W}_{l,i}\approx\exp[-\sum_{j=1}^{R} \varepsilon_{r,j}\Gamma_j'(\sum_{m=1}^Sq_m\Omega(1-P^{\text{Gen},W}_{l-1,m}))],}
\end{equation}
where $\varepsilon_{r,j}$ is the overhead contributed by relay $j$ at the destination. Unfortunately, the corresponding joint optimization problem across all relay-degree distributions is intractable. However, the density evolution expression can further be simplified by assuming the same relay-degree distribution at all the relays, $\Gamma_1(x)=\Gamma_2(x)=\ldots=\Gamma_R(x)=\Gamma(x)$ as
\begin{equation}\label{GUEP2}
P^{\text{Gen},W}_{l,i}\approx\exp[-\Gamma'(\sum_{m=1}^S q_m\Omega(1-P^{\text{Gen},W}_{l-1,m}))\sum_{j=1}^{R}\varepsilon_{r,j}].
\end{equation}
\begin{figure}
  \begin{center}
    \includegraphics[width=65mm]{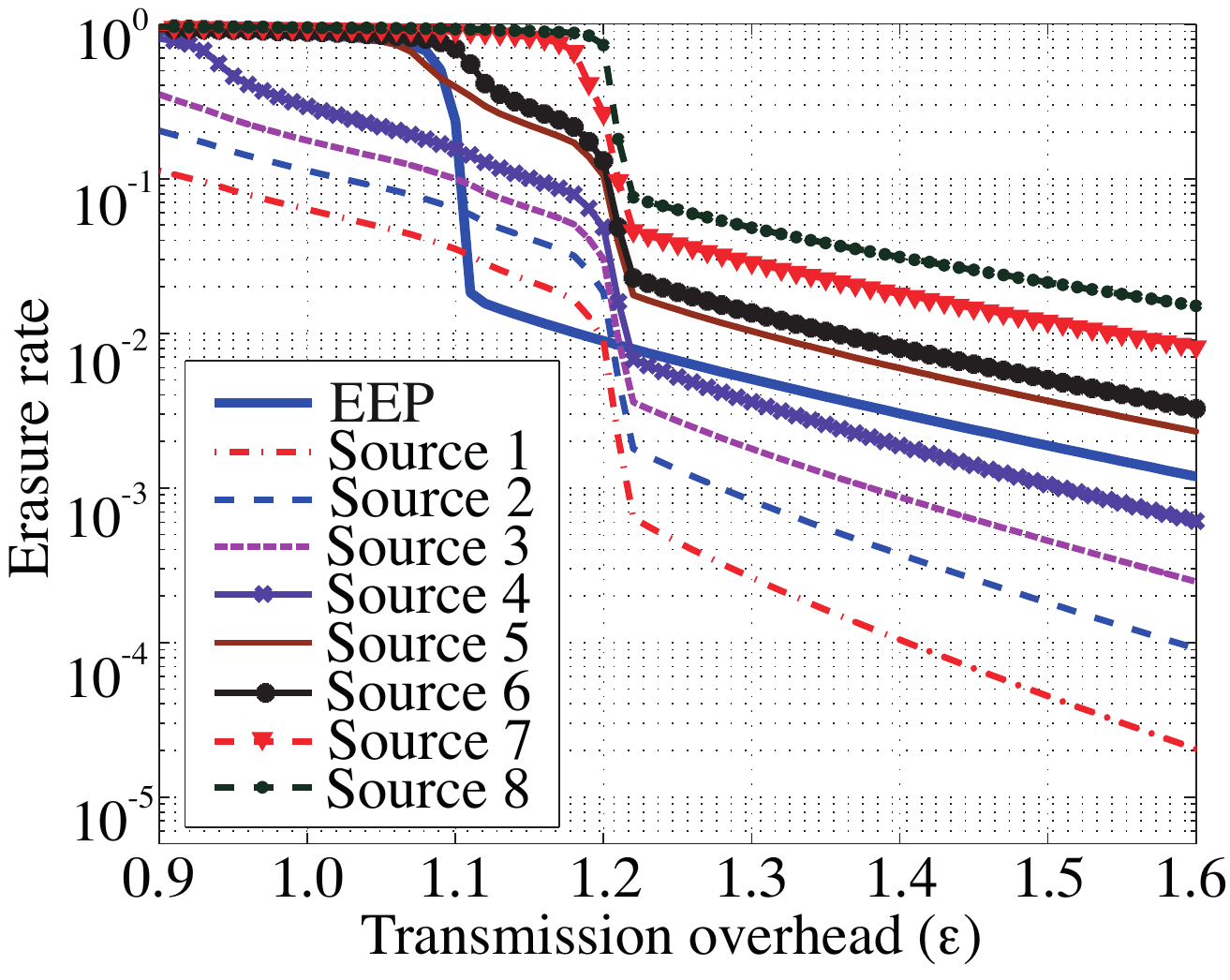}
    \caption{UEP of eight-sources, three-relays DLT code. }\label{Th_UEP_S8_R3}
  \end{center}
\end{figure}
Interestingly, for this case the decoding graph formed at the decoder is similar to the single-relay network. Therefore we can straightforwardly use the linear programs LP1 and LP2 to obtain optimized relay-degree distribution for EEP and UEP performance, respectively. This optimized relay-degree distribution is then used at each relay.

As an example, the asymptotic performance of an eight-sources, three-relays network is shown in Fig.~\ref{Th_UEP_S8_R3} for EEP and UEP performances. Each relay uses the relay-degree distribution obtained from LP1 for an eight-sources scenario and only one relay randomly selected is allowed to transmit in the relay transmission phase. We further assume that $\alpha_1=0.08$, $\alpha_2=\alpha_3=0.10$, $\alpha_4=0.13$, $\alpha_5=0.12$, $\alpha_6=\alpha_7=0.15$ and  $\alpha_8=0.17$. The erasure probabilities between relays and destination are $\delta_{1d}=0.1$, $\delta_{2d}=0.08$ and $\delta_{3d}=0.05$, respectively. The source selection probability function is specified as
\begin{eqnarray}\label{}
  q(x)\hspace{-1mm}&=&\hspace{-1mm}0.130x+0.140x^2+0.125x^3+0.145x^4 \nonumber\\
  \hspace{-1mm}&&\hspace{-1mm}+0.110x^5+0.130x^6+0.110x^7+0.110x^8,
\end{eqnarray}
to ensure source $i$ is more protected than source $j$ for $i<j$. In Fig.~\ref{Th_UEP_S8_R3} we observe that sources $1$ to $4$ are more protected than the EEP level, while sources $5$ to $8$ are less protected than the EEP level.

The proposed DEWLT codes can similarly be applied, leading to the corresponding density evolution expression as
\begin{align}\label{GDEWLT1}
P^{\text{Gen},EW}_{l,i}&\approx \exp[-\sum_{j=1}^R \varepsilon_{rj}\sum_{j=i}^I\frac{1}{\sum_{t=1}^i\Pi_t}\nonumber\\
&\hspace{3mm} \times\Gamma_{jW}'(\sum_{m=1}^S q_m\Omega(1-P^{\text{Gen},EW}_{m,l-1}))].
\end{align}
For $\Gamma_{1W}(x)=\Gamma_{2W}(x)=\ldots=\Gamma_{RW}(x)=\Gamma_{W}$, the density evolution expression can further be simplified as
\begin{align}\label{GDEWLT2}
P^{\text{Gen},EW}_{l,i}& \approx\exp[-\Gamma_{W}'(\sum_{m=1}^S q_m\Omega(1-P^{\text{Gen},EW}_{m,l-1}))\nonumber \\
&\hspace{3mm}\times \sum_{j=1}^R \varepsilon_{rj}\sum_{j=i}^I\frac{1}{\hspace{-1.5 mm}\sum_{t=1}^i\Pi_t}].
\end{align}

For the case where the number of importance classes $I$ and the number of relays $R$ are the same, the DEWLT codes can be applied in a different way.
Here, each importance class, and thus each corresponding relay-encoding window is assigned to different relays, e.g., window $\varpi_j$ is assigned to relay $j$ using relay-degree distribution $\Gamma_{jW}(x)$. The corresponding density evolution expression for an $R$-relay network is given by
\begin{equation}\label{GEWDLT}
P^{\text{Gen},EW}_{l,i}\approx\exp[- \sum_{j=i}^R \frac{\varepsilon_{r,j}}{\sum_{t=1}^i\Pi_t}\Gamma_{jW}'(\sum_{m=1}^S\Omega(1-P^{\text{Gen},EW}_{l-1,m}))].
\end{equation}


\section{Lossy Source-Relay links}
\label{Sec:UEP_lossy_single}
We consider a network with $S$ sources and one relay for the case with lossy source-relay links, as this scenario contains all relevant design challenges. Now, the assumption of lossless source-relay links is replaced with the assumption that the links between sources $i$, $i=1, 2, ..., S$, and the relay are BECs with erasure probabilities $\delta_{i}$. The link between the relay and the destination is still a BEC with erasure probability $\delta$.

Here, we first consider the use of conventional DLT codes. A one-bit buffer is then introduced to ensure low-delay encoding, and to effectively eliminate the performance loss caused by lossy source-relay links. We further improve the coding scheme and achieve UEP by applying the cancellation-avoiding encoding process detailed in Algorithm~\ref{alg2}, as well as the $D$-bits buffer-based relay-encoding strategy introduced in Subsection \ref{Subsec:EEP_lossless_single}. To cater with lost packets and avoid extra redundancy in $D$-bits buffer-based DLT codes, we also introduce Algorithm~\ref{alg3} specifically for lossy source-relay links.

\subsection{Conventional DLT codes}
The conventional DLT coding scheme proposed in \cite{SPD09} is developed for lossless source-relay links. When applied in a network with lossy source-relay links, a relay-coded bit in transmission round $n$ is determined as
\begin{equation}\label{}
z^n=\overset{S}{\underset{i=1}{\bigoplus}}\zeta_i\beta_i c_{i}^{n},
\end{equation}
where $\beta_i=1$ if the transmission of coded bit $c_{i}^{n}$ from source $i$ is received at the relay, and $\beta_i=0$ otherwise. Applying the scheme for lossy source-relay links, the relay may not be able to conduct the required encoding process unless all the selected coded bits are received from the respective sources. One alternative is to select only from current received bits which results in inferior coding. Otherwise, the relay will defer transmission to the next transmission round. It follows that the encoding process is subject to a random delay, which may be impractical for low-latency applications. To overcome this problem, we propose two buffer-based DLT strategies that eliminate the detrimental effects caused by the lossy source-relay links.


\subsection{One-Bit Buffer-Based DLT Codes}
Introducing a one-bit buffer for each source-relay link at the relay, the coded bit from source $i$ is stored in buffer $B_i$ for $i=1,2,...,S$. In every transmission round each buffer is updated with the reception of a new coded bit from the corresponding source. Due to the lossy source-relay links, there will be an initial delay before all the source buffers at the relay are filled. If a coded bit from source $i$ is erased the relay will use the previously stored coded bit at buffer $B_i$. Therefore, once the buffers are all loaded, the transmission rounds can run continuously since coded bits from all sources are instantaneously available at the relay. The initial delay is $\Delta$ rounds which depends on the largest value of $\delta_i$. Typically the initial delay is negligible as $\Delta \ll N$ due to the small values of $\delta_{i}$ for $i=1,2,\ldots,S$. The relay-coded bit in round $n$ is determined as
\begin{equation}\label{}
z^n=\overset{S}{\underset{i=1}{\bigoplus}}\psi_i b_i,
\end{equation}
where $b_i$ is the coded bit in buffer $B_i$. The one-bit buffer-based DLT code is well suited for low-latency applications. The relay selects at most one coded bit from each source. Thus, no modification is required for the encoding process at the sources, where a conventional LT code can be used. The introduction of simple one-bit buffers virtually converts the lossy source-relay links to lossless links with respect to the relay encoding process. Thus, the asymptotic performance with lossy source-relay links is similar to the lossless source-relay scenario described in \eqref{eqn:DEEEP} for EEP. Consequently we can use LP1 to obtain the optimized relay-degree distribution for EEP performance.


\subsection{$D$-Bit Buffer-Based DLT Codes}
\label{Sec:Dbuf_lossy_single}
With the introduction of a one-bit buffer, the resulting codes face similar drawbacks as the conventional DLT codes, such as a high erasure floor for a small number of sources and incapability to accommodate UEP requirements. Moreover, we cannot achieve EEP performance for sources having different information block sizes. An obvious solution is to apply the $D$-bits buffer-based relay-encoding strategy introduced in Subsection \ref{Subsec:EEP_lossless_single}. However, we cannot apply the combining process detailed in Algorithm~\ref{alg1}. The self-cancellation cannot be avoided due to the loss of coded bits in the source transmission round. As a consequence, an arbitrary buffer may have more than one bit from the same class. Therefore we introduce a new combining scheme at the relay for the lossy source transmission round.

In round $n$ each newly received coded bit $c_{i}^{n}$ from source $i$ is stored in a specific position in buffer $B_i$. The newly received encoded bit $c_{i}^{n}$ is always stored in location $v_l=\mod(n-1,D)+1$ in buffer $B_i$ in round $n$. This ensures that each location in the buffer have coded bits from different classes at the sources. The newly received  coded bit overwrites the previously stored bit in $B_i$ for $i=1, 2,\ldots, S$. In case $c_{i}^{n}$ is erased, no overwrite operation is performed and the previously stored buffer bit is used in the combining process at the relay. Again, there will be an initial delay of $(D+\Delta)$ rounds before all the source buffers at the relay are filled, after which the relay can transmit continuously in every transmission round. It should be noted that the initial delay is higher than its lossless counterparts due to the packet loss in the source transmission round. However the initial delay required for loading the buffers is negligible as $(D+\Delta) \ll N$ and $\delta_i$ is typically small for $i=1,2,\ldots,S$. The selection within the buffer can be performed randomly or sequentially. However, we have observed that these selection approaches within the buffers have degraded performance. For improved performance, high priority should be given to the newly arrived coded bit to avoid extra redundancy at the destination. For this purpose, the relay-coded bit in transmission round $n$ is determined as
\begin{equation}\label{Rbit_lossy}
z^n=\overset{S}{\underset{i=1}{\bigoplus}} \psi_i\left(\overset{d_i}{\underset{m=1}{\bigoplus}} b^{u_m}_{i}\right),
\end{equation}
where $u_m=\mod(v_l-m+D,D)+1$. The combining process at the relay is detailed in Algorithm~\ref{alg3}. Similar to the one-bit buffer, we effectively convert the lossy source-relay links to lossless links, and as before $d_{\Gamma,\text{max}}$ is not limited by the number of sources present in the network. Therefore the asymptotic performance over lossy source-relay links is virtually the same as the asymptotic performance of the corresponding lossless counterparts.

\begin{algorithm}[t]                                               
\caption{: Proposed Combining Scheme at Relay}                     
\label{alg3}                                                       
\begin{algorithmic}
\STATE {\bf Initialization:} Wait $D+\Delta$ transmission rounds until all buffers are filled.
\WHILE{unless relay receives an acknowledgment from the destination to halt transmission}
\STATE {\bf Step 1: } After completion of source transmission phase in round $n$, update those buffers with overwrite operation which have received their corresponding source-encoded bits. The source encoded bit $c_{i}^{n}$  will be stored at $v_l=\mod(n-1,D)+1$ in $B_i$ for all $i$ with erasure free reception.
\STATE {\bf Step 2: } Sample $\Gamma(x)$ to determine the number buffer bits $d$ to be combined to form the relay-coded bit $z^n$.
\STATE {\bf Step 3: } Sample $q(x)$ to determine the buffers and number of bits from each buffer i.e. $d_i$ for $i=1,2,\ldots,S$.
\STATE {\bf Step 4: } Determine $z^n=\overset{S}{\underset{i=1}{\bigoplus}} \psi_i\left(\overset{d_i}{\underset{m=1}{\bigoplus}} b^{u_m}_{i}\right)$.
\STATE {\bf Step 5: } Update the generator matrix at the relay.
\STATE {\bf Step 6: } Transmit the coded bit $z^n$ over relay-destination link.
\ENDWHILE

\end{algorithmic}
\end{algorithm}


\subsection{Extensions to UEP and $S$-Sources, $R$-relays Network}
Following the approach in Subsection \ref{Sec:UEP_lossless_single}, the UEP capabilities are determined by the information block sizes of the sources, and the source-selection probability distribution described by $q(x)$. The corresponding values of $w_i$, $i=1, 2, ..., S$, determine the level of UEP between the sources. The encoding process at the sources and the combining process at the relay remain the same as discussed in Subsection \ref{Sec:Dbuf_lossy_single}. As above, the introduction of buffers at the relay converts the lossy source-relay links to virtually lossless links, and thus we can directly apply the density evolution expressions in \eqref{DEUEP}. We can further obtain optimized degree distribution for UEP using the linear program outlined in LP2. The extension to a general $S$-sources, $R$-relays network is straightforward, and therefore not described any further.

Potential drawbacks of our proposed DLT coding scheme are that the relay complexity is higher than for the conventional DLT coding scheme, both in terms of memory requirements and computations. The additional memory requirements is obviously due to the introduction of buffers, while
\begin{figure}
  \begin{center}
    \includegraphics[width=65mm]{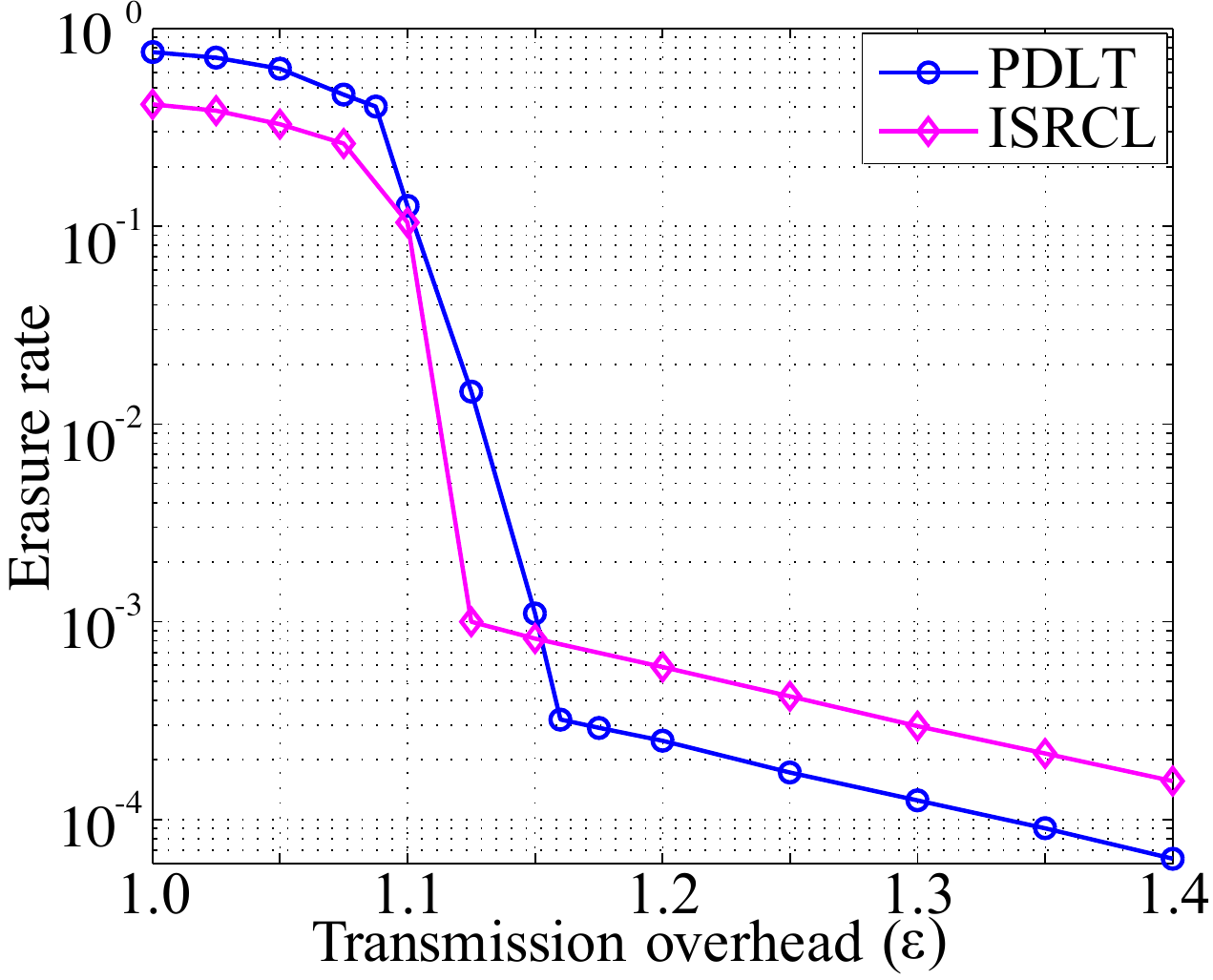}
    \caption{\textcolor{red}{EEP comparison of the ISLRC scheme in \cite{LiauKimYousefi13TCOM} and the proposed DLT codes. Each source uses a RSD with $K=100$, $c=0.05$, and $\delta=0.5$.}}\label{Sim_EEP_S2_R1}
  \end{center}
\end{figure}
the increase in computational complexity is attributed to storing and retrieving data from the buffers, as well as the maximum number of source-encoded bit to combine is significantly higher. Thus, the superior performance for our proposed DLT codes is achieved at the expense of extra memory and higher computational complexity at the relay as compared to the conventional DLT coding scheme. However due to the small memory buffers, our buffer-based DLT codes are still attractive in resource-limited relay applications similar to \cite{LiauKimYousefi13TCOM}.


\section{Numerical Results}
\label{Sec:Numerical}
In this section we consider a series of examples for numerical evaluation, and comparison to theoretical results obtained through density evolution and performance bounding. We first consider the case of lossless source-relay links, before discussing examples for lossy source-relay links.

\subsection{Lossless Source-Relay Links}
We first compare the performance of our proposed DLT (PDLT) codes to the performance of the improved Soliton-like rateless coding (ISLRC) scheme outlined in \cite{LiauKimYousefi13TCOM} for $K=10000$ over a network with two sources and a single relay. The relay parameters for ISLRC scheme are $M=10$, $\Lambda=0.90$ and $\zeta=0.50$ 
\begin{figure}
  \begin{center}
    \includegraphics[width=65mm]{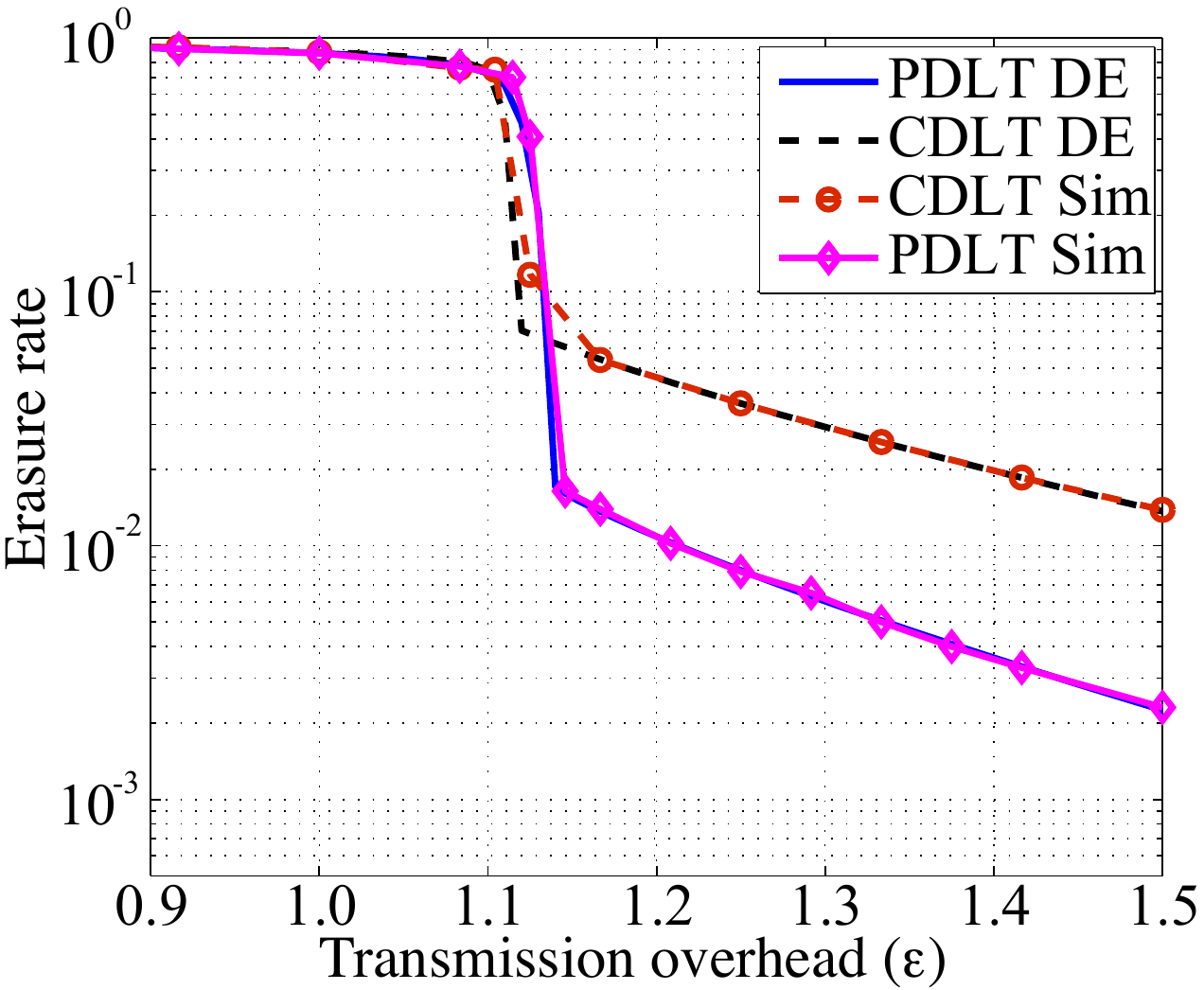}
    \caption{EEP comparison of conventional and proposed DLT codes with $K=24000$. }\label{Sim_EEP_S4_R1}
  \end{center}
\end{figure}
(see \cite{LiauKimYousefi13TCOM} for details). The relay degree distribution for our DLT coding scheme is obtained from LP1 with $d_{\Gamma,\text{max}}=4$. The performance comparison is shown in Fig.~\ref{Sim_EEP_S2_R1} for a lossless relay-destination link and equal information block size sources. The improved performance of our proposed DLT coding scheme in terms of erasure floor can easily be observed in Fig.~\ref{Sim_EEP_S2_R1} as compared to the ISLRC scheme for the given parameters. The abrupt improvement in the performance of the proposed DLT code as shown in Fig.~\ref{Sim_EEP_S2_R1} also confirms the the well-known avalanche effect in LT decoding \cite{Mackay05}.

\begin{figure}
  \begin{center}
    \includegraphics[width=65mm]{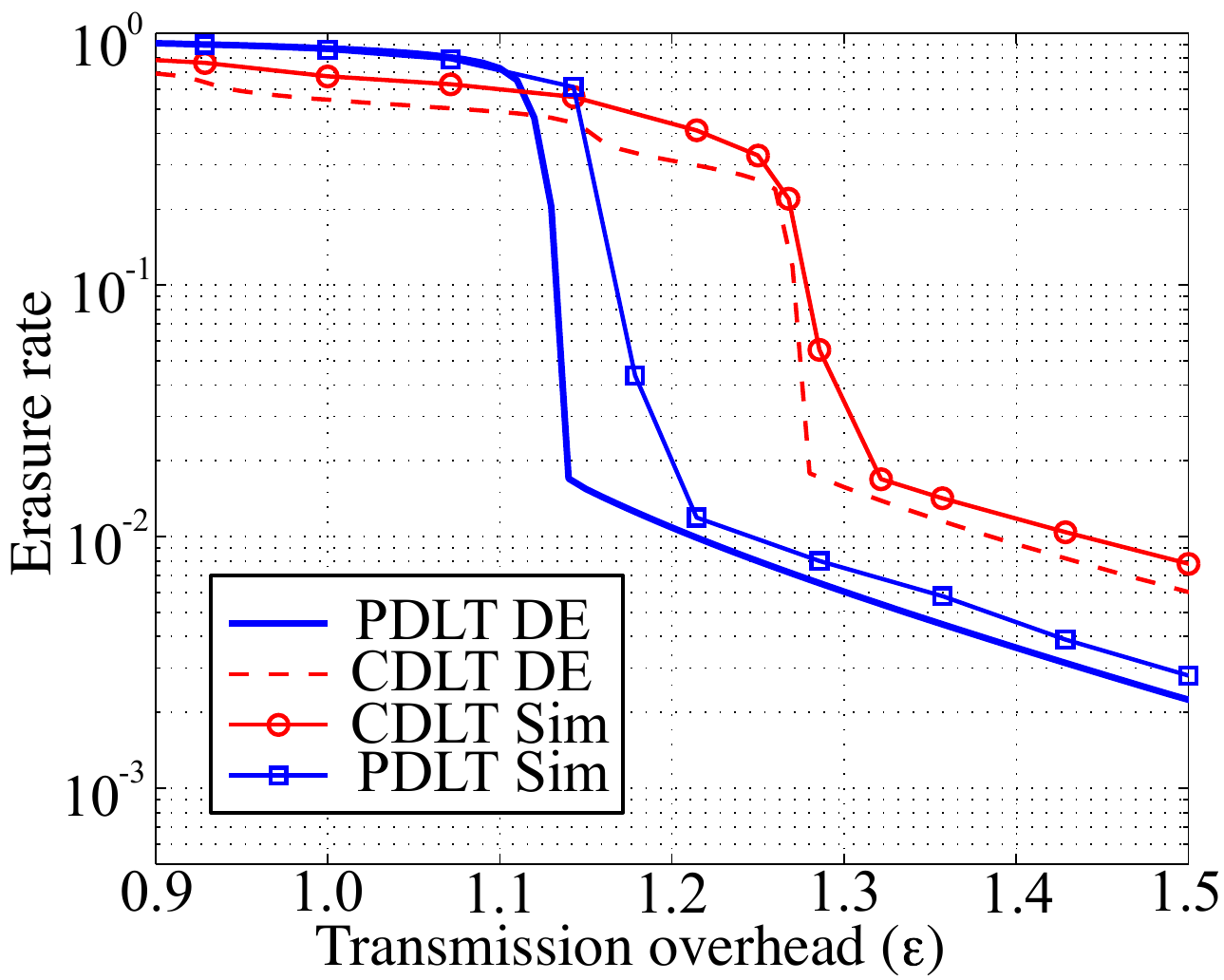}
    \caption{EEP of conventional and proposed DLT codes with $K=28000$ and different block sizes. }\label{Sim_EEP_differentsizes}
  \end{center}
\end{figure}
We subsequently consider the EEP performance of conventional DLT (CDLT) codes as compared to the EEP performance of our proposed DLT codes over a network with four sources and a single relay.
To emphasize the performance gain obtained by enabling a higher maximum degree $d_{\Gamma,\text{max}}$ of the relay degree distribution due to buffers in our proposed DLT codes, we assume each source has equal information block size. The source-relay links are lossless, and the erasure probability between the relay and the destination is $\delta=0.1$. For the conventional DLT code $ d_{\Gamma,\text{max}}=S=4$ and for our proposed DLT codes $d_{\Gamma,\text{max}}=D=8$, while the corresponding relay degree distributions are obtained from LP1. The results are presented in Fig.~\ref{Sim_EEP_S4_R1}, where we can easily observe the improved performance of our proposed scheme as compared to the conventional DLT coding scheme.
To highlight the performance improvement for different information block sizes, the example is extended to eight sources with different information bit block sizes such that $\alpha_1=\alpha_2=5/28$, $\alpha_3=\alpha_4=1/7$, $\alpha_5=\alpha_6=3/28$, and $\alpha_7=\alpha_8=1/14$. From Fig.~\ref{Sim_EEP_differentsizes} it is clear that our proposed DLT coding scheme provides significantly better performance as compared to the conventional DLT coding scheme. The degree distribution for both schemes are achieved from the linear program LP1 for $d_{\Gamma,\text{max}}=8$. For reference, the corresponding asymptotic performances, based on density evolution (DE), are also shown in Fig.~\ref{Sim_EEP_differentsizes}.

\begin{figure}
  \begin{center}
    \includegraphics[width=65mm]{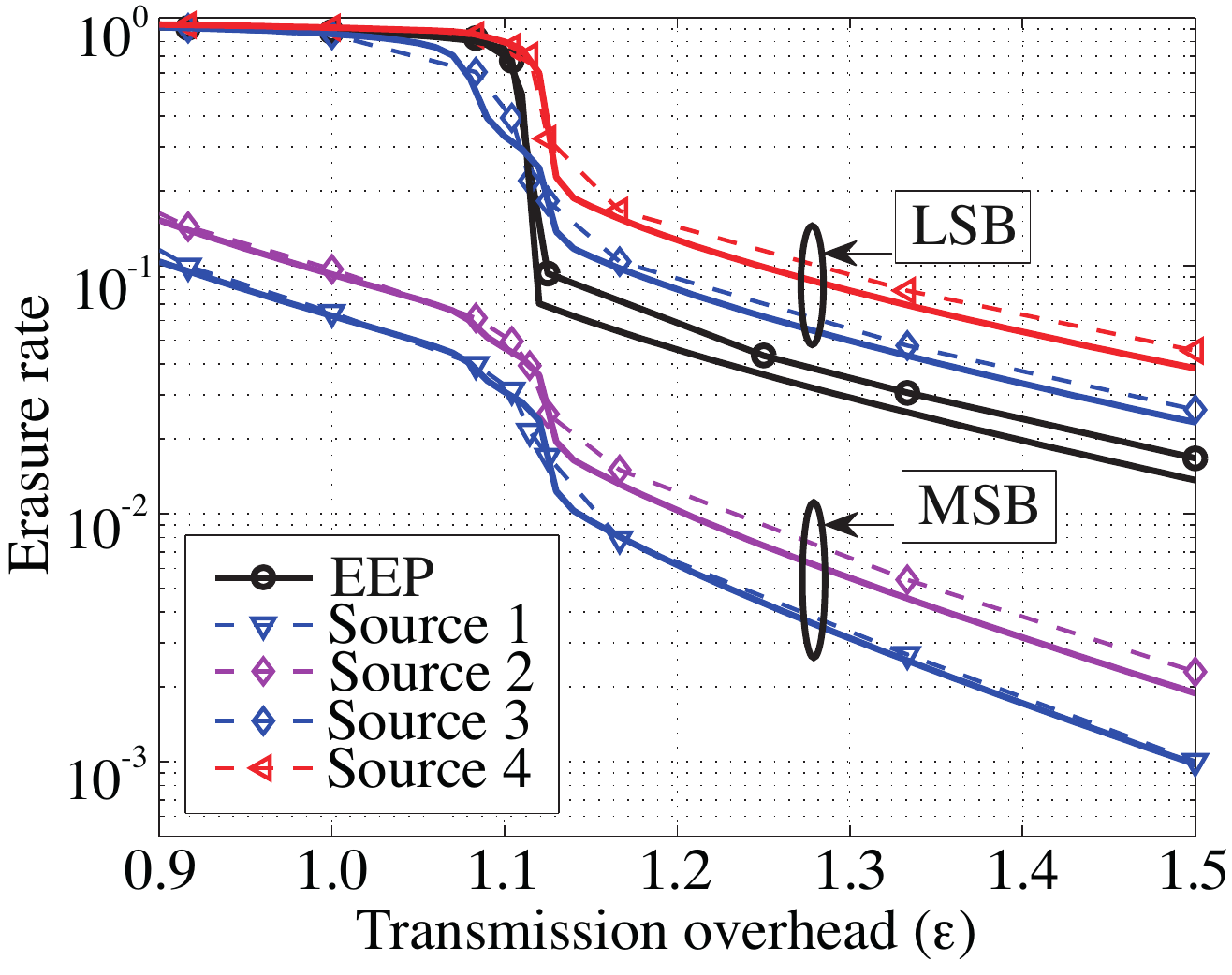}
    \caption{UEP of four-sources DLT codes with $K=24000$, EEP optimized relay-degree distribution. }\label{Sim_UEP_S4_R1_1}
  \end{center}
\end{figure}
The UEP performance for a four-sources DLT code is shown in Fig.~\ref{Sim_UEP_S4_R1_1} with dashed lines, where we assume the same parameters as used for the example in Fig.~\ref{Th_UEP_S4_R1_Lossless_1}. Here sources $1$ and $2$ are significantly better protected than sources $3$ and $4$. For reference the EEP performance of the proposed DLT code is also given in Fig.~\ref{Th_UEP_S4_R1_Lossless_1}. Moreover we have included the corresponding asymptotic performances as solid lines.

\begin{figure}
  \begin{center}
    \includegraphics[width=65mm]{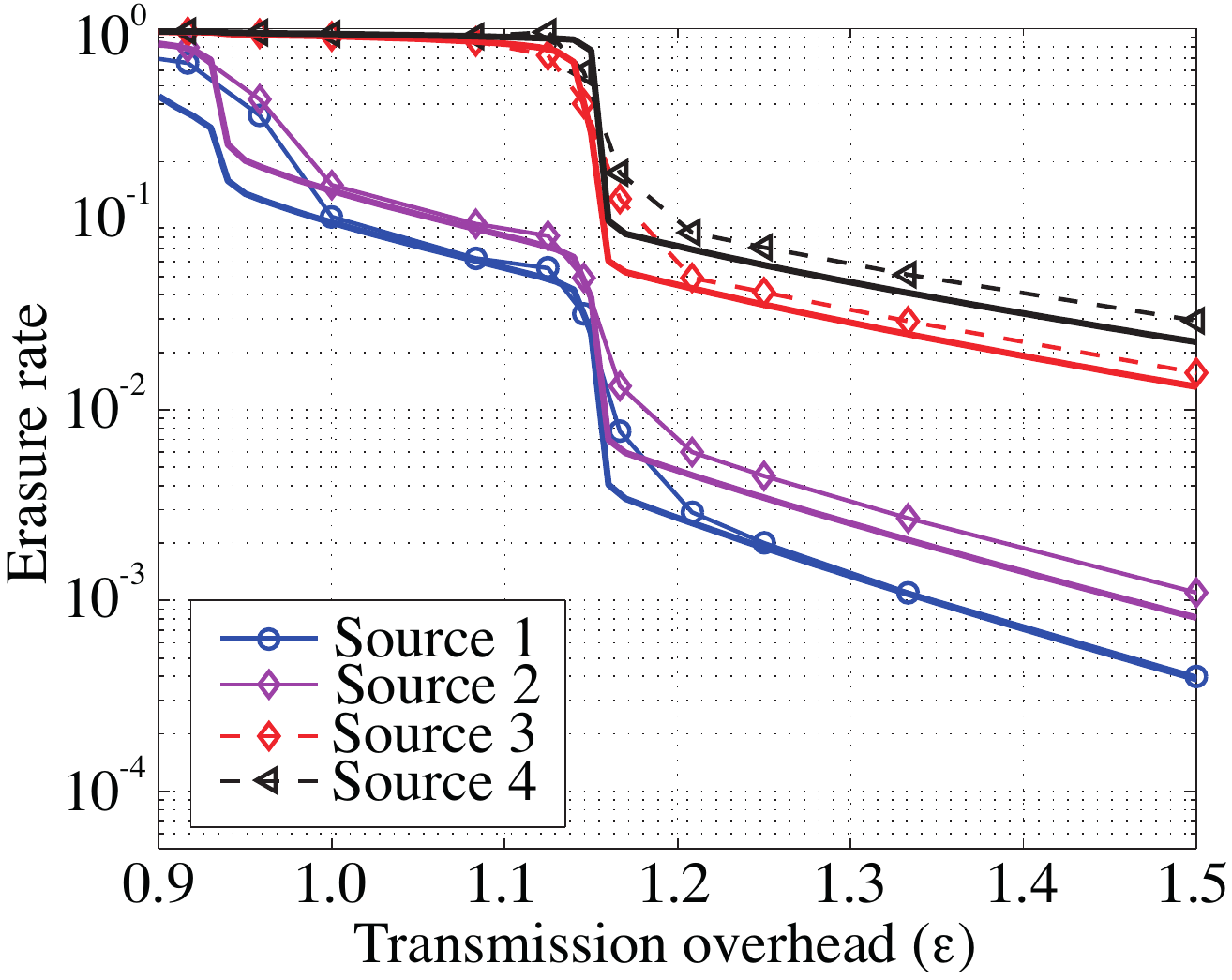}
    \caption{UEP of four-sources DLT codes with $K=24000$, UEP optimized relay-degree distribution. }\label{Sim_UEP_S4_R1_2}
  \end{center}
\end{figure}

The UEP performance for the same parameters,  but using the degree distribution obtained from LP2, is shown in Fig.~\ref{Sim_UEP_S4_R1_2}. Again the corresponding asymptotic
performances are drawn as solid lines. Compared to the UEP performance based on the EEP degree distribution in Fig.~\ref{Sim_UEP_S4_R1_1}, it is clear that our optimized degree distribution has improved performance for all sources.

\begin{figure}
  \begin{center}
    \includegraphics[width=65mm]{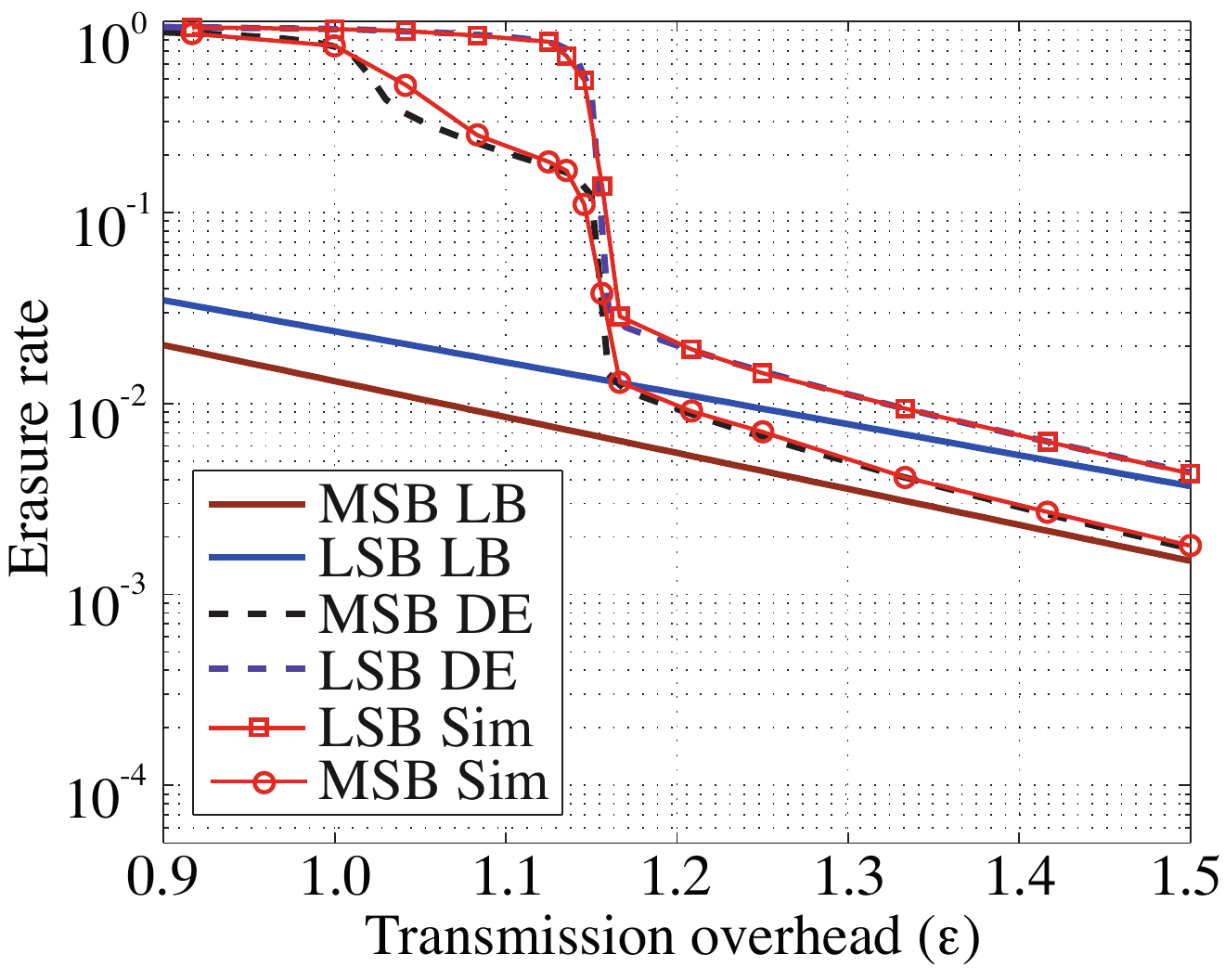}
    \caption{UEP eight-sources DEWLT codes, $K=24000$, $\alpha_i=1/8$. }\label{DEWLT_Sim}
  \end{center}
\end{figure}

The asymptotic performance through density evolution, lower bounds (LB) and the corresponding numerical results are shown for an eight-sources and two importance-classes scenario using DEWLT codes in
Fig.~\ref{DEWLT_Sim}. For a clearer view the performance of the sources $1, 2, 3, 4$ are averaged in the MSB class while
the performance of the sources $5$ to $8$ are averaged in the LSB class. For $\Gamma_{1W}$, we use the relay-degree distribution detailed in \eqref{rDD}, and for $\Gamma_{2W}(x)$, we determine the relay-degree distribution using LP1. From Fig.~\ref{DEWLT_Sim}, we conclude that the numerical results closely follow the density evolution analysis and the lower bounds.

The numerical results for an eight-sources and three-relays network is shown in Fig.~\ref{Sim_S8_R3_Lossless}. The scenario parameters are the same as for the example in Fig.~\ref{Th_UEP_S8_R3}. From the class selection probabilities and relay degree distributions, we ensure that source $i$ is more protected than source $j$ for $i<j$. Moreover, the first four sources are more protected than its EEP counterpart as demonstrated in Fig.~\ref{Sim_S8_R3_Lossless}. For reference the corresponding density evolution results from Fig.~\ref{Th_UEP_S8_R3} are also given as solid lines in Fig.~\ref{Sim_S8_R3_Lossless}.


\subsection{Lossy Source-Relay Links}

\begin{figure}
  \begin{center}
    \includegraphics[width=65mm]{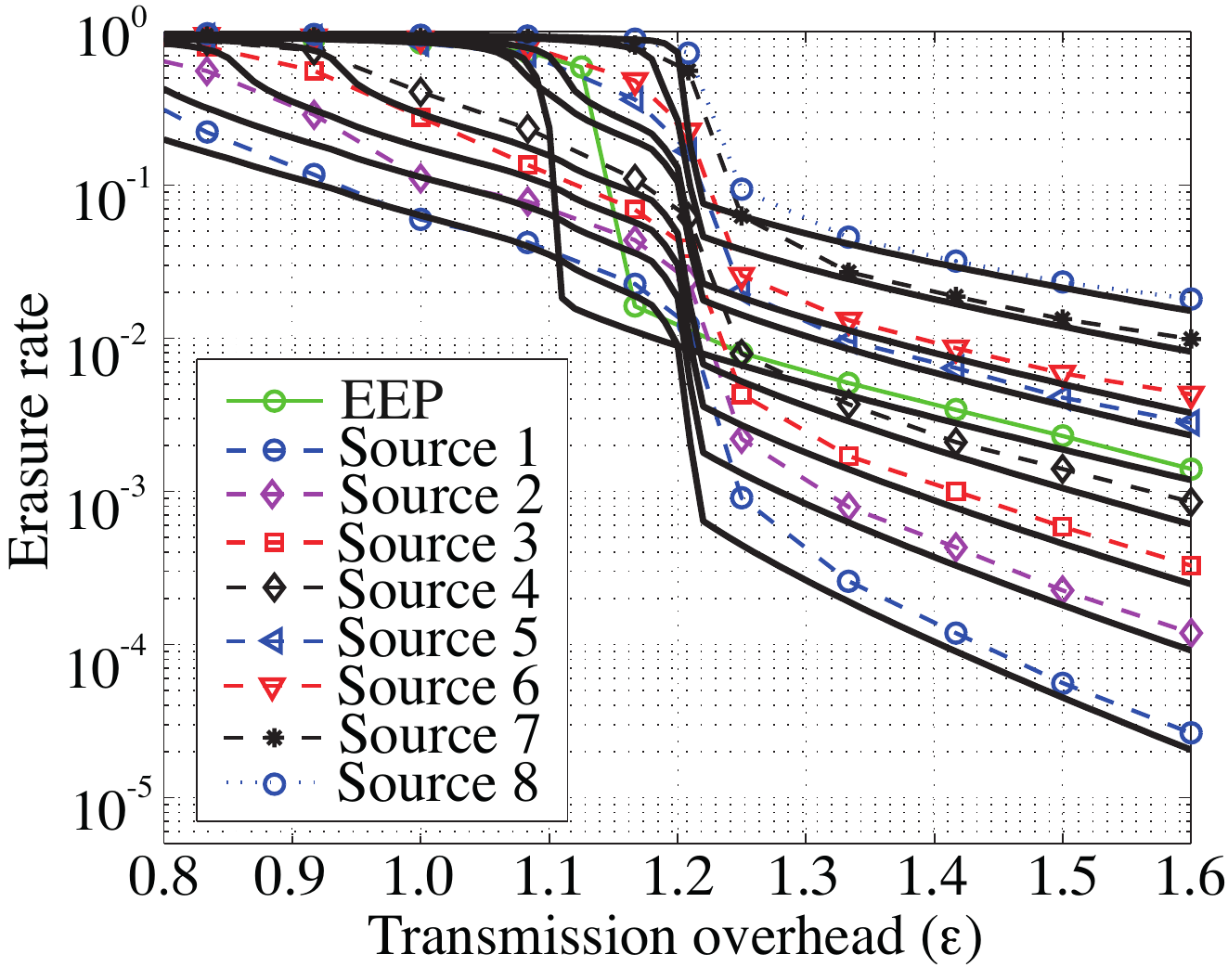}
    \caption{UEP eight-sources and three-relays network for $K=24000$. }\label{Sim_S8_R3_Lossless}
  \end{center}
\end{figure}
\begin{figure}
  \begin{center}
    \includegraphics[width=65mm]{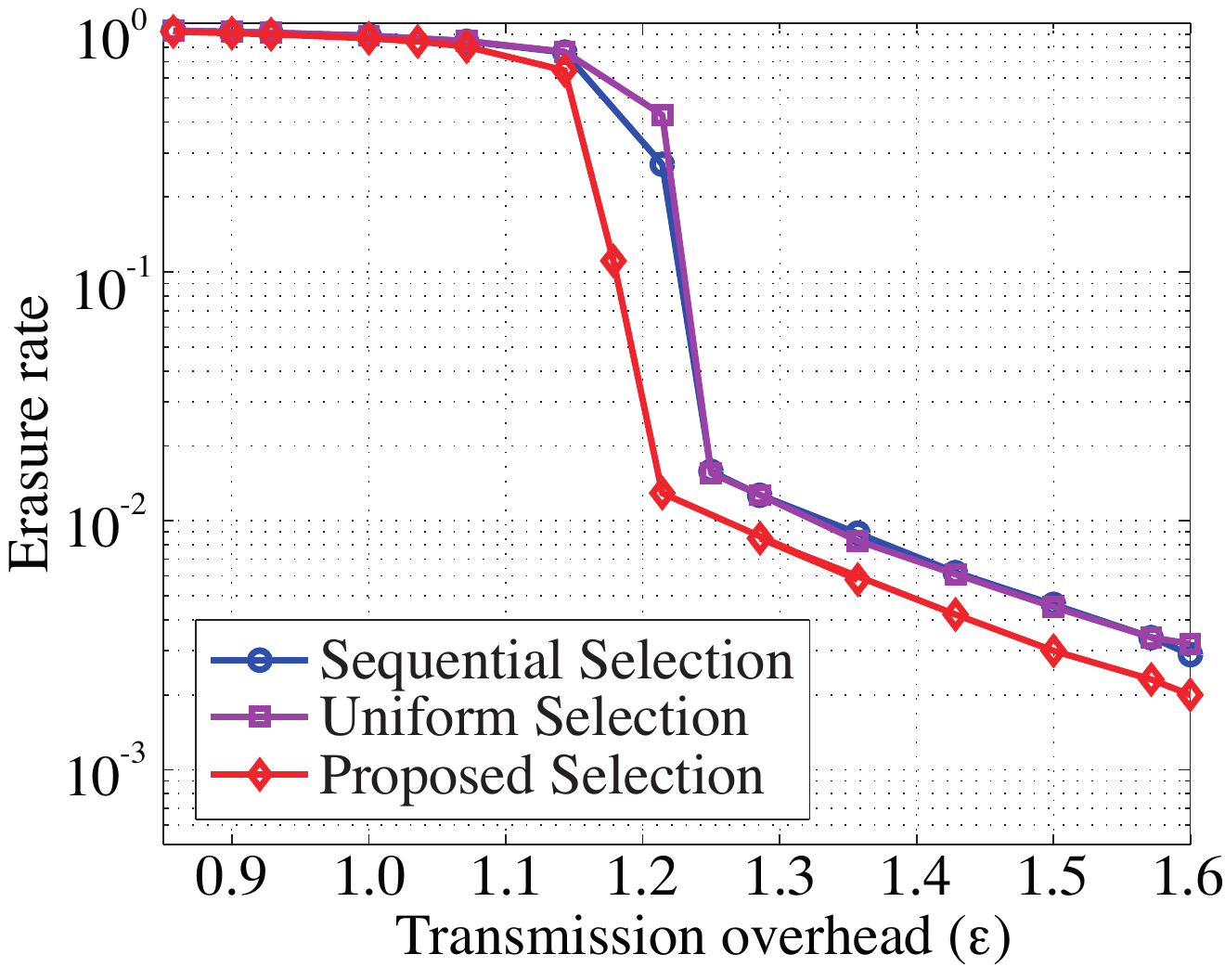}
    \caption{Performance of an eight-sources single-relay network with different selection schemes.} \label{Sim_S8_R1_differentSelection}
  \end{center}
\end{figure}

To demonstrate the effectiveness of our proposed Algorithm~\ref{alg3} for lossy links, we compare the performance of our proposed DLT codes for a network with eight sources and one  relay with DLT codes where the relay selects the bits within the buffer sequentially or uniformly-at-random. We assume the same parameters as used for the example in
Fig.~\ref{Sim_EEP_differentsizes} with additional lossy source-relay channels such that $\delta_{i}=0.05$ for all $i$. The corresponding performance improvement of Algorithm~\ref{alg3} as compared to other selection approaches is presented in Fig.~\ref{Sim_S8_R1_differentSelection}.

\begin{figure}
\begin{center}
    \includegraphics[width=65mm]{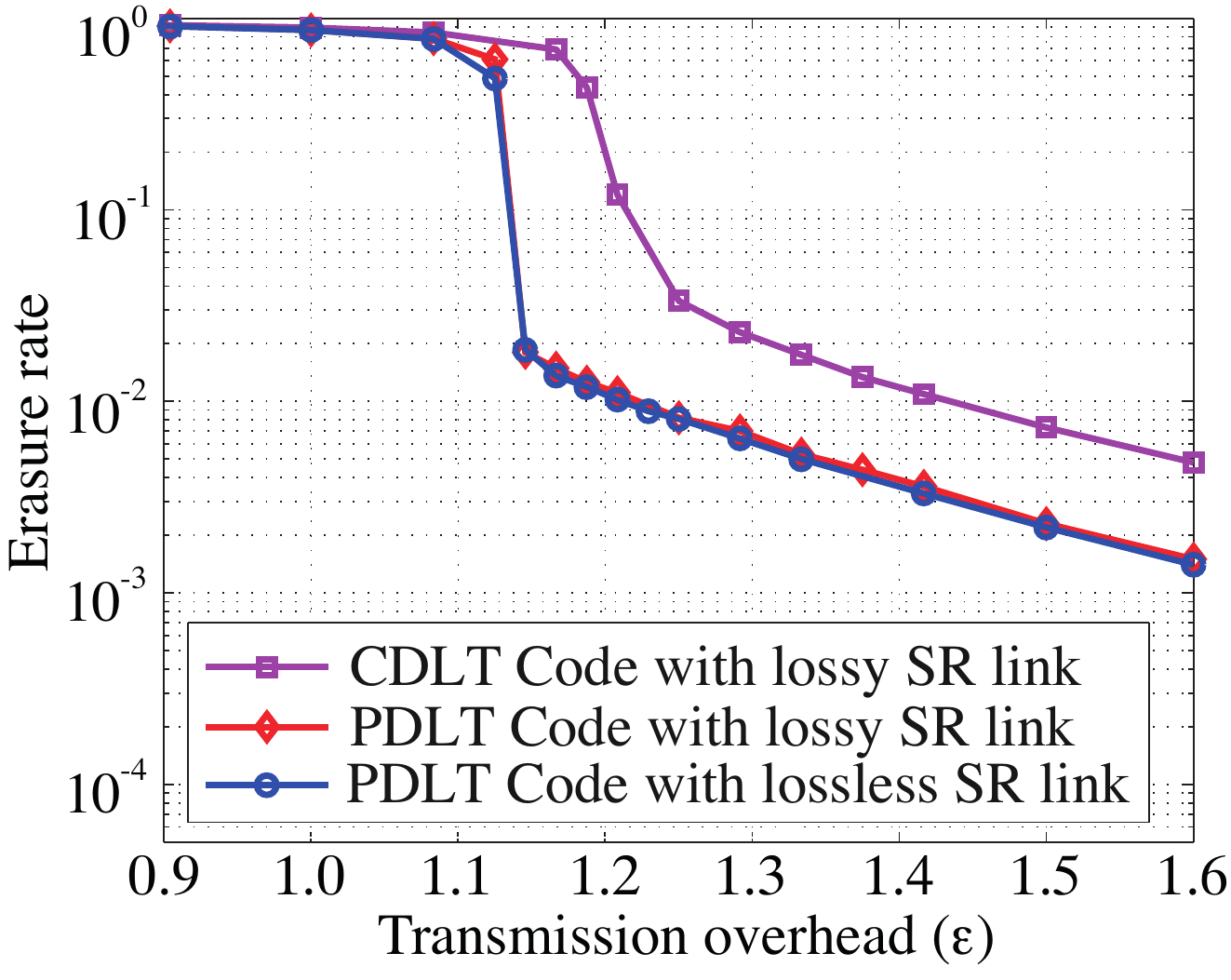}
    \caption{EEP performance over lossy source-relay links for an eight-sources single-relay network.} \label{Sim_S8_R1_Lossless_vs_Lossy}
  \end{center}
\end{figure}

The performance of the proposed one-bit buffer-based DLT coding scheme for a single-relay network with eight lossy source-relay links is shown in  Fig.~\ref{Sim_S8_R1_Lossless_vs_Lossy}. The parameters for this set up are $\delta=0.1$ and $\delta_{i}=0.05$ for all $i$, while the relay uses the degree distribution obtained from LP1. To avoid a performance loss for the conventional DLT coding scheme due to different information block sizes, we consider sources of equal sizes. We also consider the corresponding lossless source-relay (SR) scenario in Fig.~\ref{Sim_S8_R1_Lossless_vs_Lossy} for
reference. Interestingly, we observe that our proposed DLT codes for lossy source-relay links have almost the same performance as that for lossless source-relay links.

\begin{figure}
  \begin{center}
    \includegraphics[width=65mm]{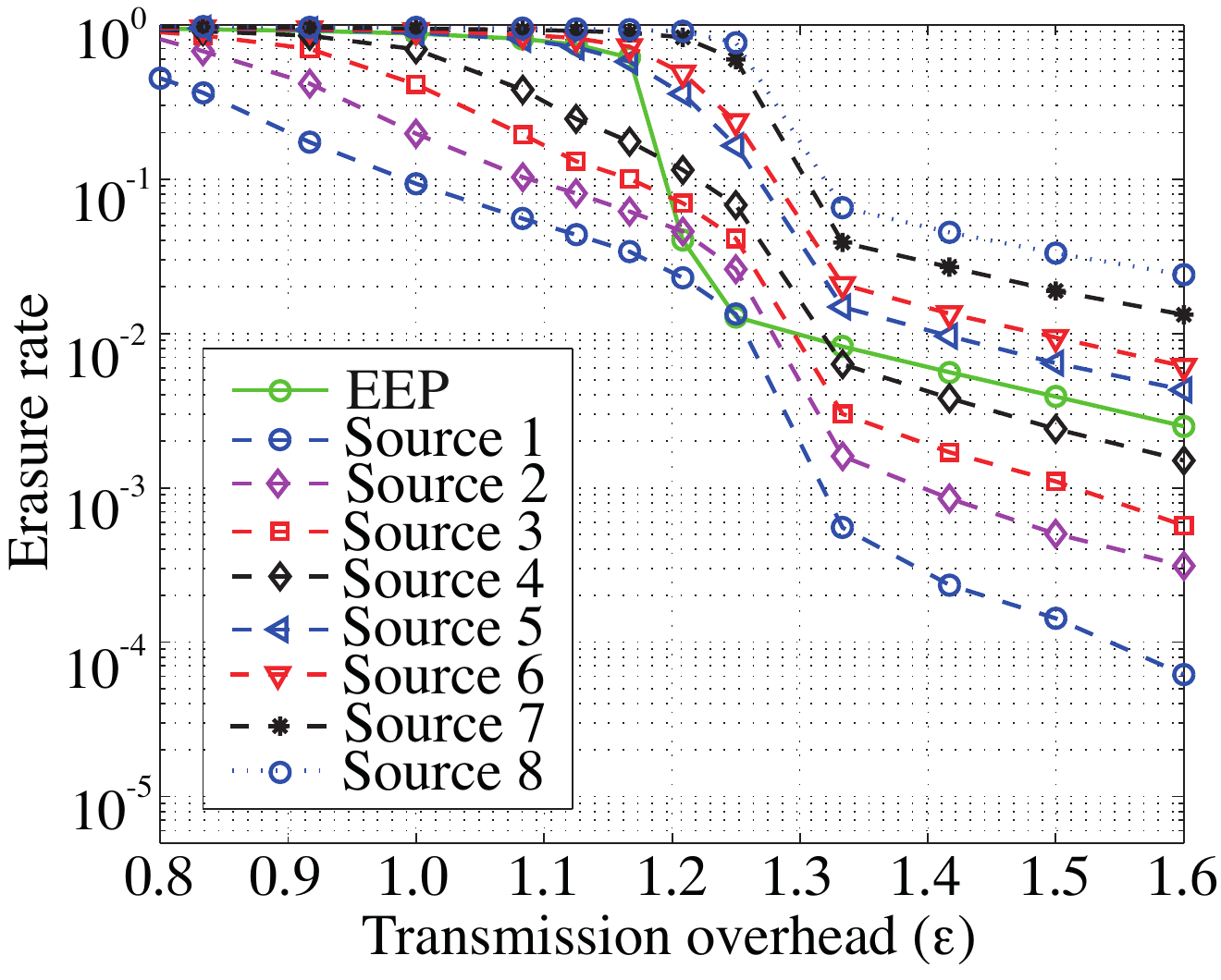}
    \caption{UEP eight-sources, three-relays, lossy source-relay links. }\label{Sim_S8_R3_Lossy}
  \end{center}
\end{figure}

The UEP performance of an eight-sources three-relays network with our proposed $D$-bit buffer-based DLT code is shown in Fig.~\ref{Sim_S8_R3_Lossy}. All the parameters are the same as for Fig.~\ref{Sim_S8_R3_Lossless}, except the links between the sources and the relays are lossy with $\delta_{ij}=0.05$ for all $i$ and $j$. By comparison with Fig.~\ref{Sim_S8_R3_Lossless}, we can observe that our proposed buffer-based DLT codes achieve similar performance as their lossless counterparts. Thus, with the proposed $D$-bit buffer, we can achieve improved EEP and UEP performance over lossy links between sources and relays, as well as between relays and the destination. The approach can readily be extended to different values of channel erasure probabilities of source-relay and relay-destination links.


\section{Conclusions}
\label{Sec:Conclusion}

We investigated the design and performance of DLT codes over networks with multiple sources and multiple relays. We proposed an encoding process at the sources and a combining process at the relays for improved performance. The proposed DLT codes have improved EEP performance and unlike conventional DLT codes, provide EEP irrespective of the information block length sizes at the sources. The structure of our proposed DLT codes were then exploited for UEP performance. The proposed UEP-DLT codes exhibited UEP for networks having arbitrary number of sources and relays in contrast to its conventional counterparts. The proposed coding scheme can be used as stand-alone or combined with other schemes such as DEWLT coding scheme. Furthermore, the proposed DLT codes virtually convert lossy source-relay links to corresponding lossless links. Consequently, our proposed DLT codes is a natural choice in low-latency data transmission especially for lossy source-relay links.


\bibliographystyle{IEEEtran}
\bibliography{IEEEabrv,references_main}

\end{document}